\documentclass[aps,pra,superscriptaddress,twocolumn,longbibliography,nofootinbib]{revtex4-2}
\usepackage[utf8]{inputenc}
\usepackage[table,xcdraw]{xcolor}
\usepackage{graphicx}
\usepackage{xcolor}
\usepackage{hyperref}
\usepackage{amsthm}
\usepackage{amsmath}
\usepackage{amsfonts}
\usepackage{comment}
\usepackage{color}
\usepackage{tabularx}
\usepackage{quantikz}
 \usepackage[inkscapearea=page]{svg}
\hypersetup{hidelinks,colorlinks=true,breaklinks=true,urlcolor=blue,allcolors=blue,pdftitle={Title},pdfauthor={Author}}

\begin{document}

\theoremstyle{definition}
\newtheorem{definition}{Definition}
\theoremstyle{definition}
\newtheorem*{conjecture*}{Conjecture}

\title{Multidimensional Fourier series with quantum circuits}
\author{Berta Casas}
\email{berta.casas@bsc.es}
\affiliation{Barcelona Supercomputing Center, Plaça Eusebi G\"uell, 1-3, 08034 Barcelona, Spain.}

\author{Alba Cervera-Lierta}
\email{alba.cervera@bsc.es}
\affiliation{Barcelona Supercomputing Center, Plaça Eusebi G\"uell, 1-3, 08034 Barcelona, Spain.}

\date{\today}

\begin{abstract}

Quantum machine learning is the field that aims to integrate machine learning with quantum computation. In recent years, the field has emerged as an active research area with the potential to bring new insights to classical machine learning problems. One of the challenges in the field is to explore the expressibility of parametrized quantum circuits and their ability to be universal function approximators, as classical neural networks are. Recent works have shown that with a quantum supervised learning model, we can fit any one-dimensional Fourier series, proving their universality. However, models for multidimensional functions have not been explored in the same level of detail. In this work, we study the expressibility of various types of circuit ansatzes that generate multidimensional Fourier series. We found that, for some ansatzes, the degrees of freedom required for fitting such functions grow faster than the available degrees in the Hilbert space generated by the circuits.  For example, single-qudit models have limited power to represent arbitrary multidimensional Fourier series. Despite this, we show that we can enlarge the Hilbert space of the circuit by using more qudits or higher local dimensions to meet the degrees of freedom requirements, thus ensuring the universality of the models.

\end{abstract}

\maketitle

\section{Introduction}

Machine learning (ML) is a well-established field that aims to develop the necessary tools to extract knowledge from big data batches by drawing inferences from its patterns. The development of computational paradigms such as quantum computation opens the path to explore the use of quantum devices to perform ML tasks, which raises the question of whether quantum machine learning (QML) algorithms can offer an advantage compared to classical ones. 

QML explores the use of quantum computing devices to implement ML algorithms \cite{Biamonte2017}. In some of these algorithms, in particular, in the supervised learning ones, data stored in a classical register needs to be mapped into a quantum state to be later processed by the quantum circuit. The parameters of the circuit are optimized by minimizing a cost function that compares some expectation values obtained from the circuit with the true data labels. Several proposals exist to embed data into quantum circuits \cite{DR, Quantum_embedding, goto2021universal}. 
Particularly, the re-uploading strategy has been applied to several problems, such as classification \cite{DR} or function fitting \cite{gil2020input}. 

It has been shown that partial Fourier series emerge as the output of these models when using the re-uploading protocol \cite{Effect_data_encoding}. Fourier series are universal to represent any square-integrable function in a given interval. This result has attracted the attention of recent QML works that use classical data \cite{Heimann2022LearningCO, Shin2022ExponentialDE, Daskin2022AWT, yu2020power, gan2022fock}. Nevertheless, many applications that may require the use of QML rely on multidimensional datasets \cite{Kinga2023Quantum}. Although several works discuss the generalization of this model to multidimensional data, there needs to be a more thoughtful analysis of the implementation and scaling of the model with the data dimensions.

The current state of the art in experimental quantum computation, the so-called noisy intermediate-scale quantum (NISQ) era \cite{Preskill2018quantumcomputingin}, presents a few qubit devices with limited coherence times that impose restrictions on the circuit depth and the fidelity of quantum operations. A family of algorithms suitable for these devices are variational quantum algorithms (VQA) \cite{bharti2021noisy, VQA}, from which the aforementioned supervised QML algorithms are part of. Variational algorithms present certain noise resilience due to their hybridization with classical optimization subroutines that fine-tune the circuit parameters. However, the low number of qubits in these NISQ devices limits the size of the Hilbert space available to perform quantum computational tasks. A way to circumnavigate this issue is to increase the Hilbert space by exploiting the higher local dimensions of each quantum information unit. Indeed, all quantum systems naturally contain more than two levels, which makes translating a qubit quantum device into a qudit system with $d$ dimensions technically feasible, despite being experimentally challenging. Thus, increasing the Hilbert space from $2^n$ to $d^n$ for $n$ quantum information units might prove valuable in the near term \cite{qudit_processor, CerveraLierta2022, Erhard_2020, Castro2022, PhysRevB.95.064423, Qutrit_entanglement}.

This work analyzes the multidimensional Fourier series representation of QML circuits fed with classical data using a general formalism for qubits and beyond (qudits). We show how the quantum circuit requirements scale with the dimension of qudits and data when using the well-established data re-uploading strategy. Furthermore, we provide four types of circuit ansatzes to generate these functions and examine the constraints to ensure a proper fitting of a general series. The study of this problem sheds light on the expressibility of QML models, and in turn, provides insights into their capacity and limitations. It also addresses the question of whether these models can be considered universal approximators.

We show that, while the proposed encoding strategies generate multidimensional Fourier series, the scaling of the degrees of freedom required to fit a general function grows rapidly, in some cases faster than the Hilbert space of the Parameterized Quantum Circuit (PQC). However, we present strategies that can overcome such limitations by increasing the dimension of the local space or employing a greater number of qudits. These strategies demonstrate the capability of generating arbitrary Fourier series through re-uploading methods, providing evidence of their universality.

This work is organized as follows: In Section \ref{sec:FS with quantum circuits}, we review the formalism of the one-dimensional Fourier series, and in Section \ref{sec:Multi FS}, we extend it to the multidimensional case. Furthermore, we discuss the degrees of freedom requirements, implementations and analysis of the models in Section \ref{sec:results}. Finally, the conclusions are presented in Section \ref{sec:conclusions}.

\section{One-dimensional Fourier series with quantum circuits}\label{sec:FS with quantum circuits}

\begin{figure}[t]                         
\centering
\includegraphics[width=1\columnwidth, angle=0]{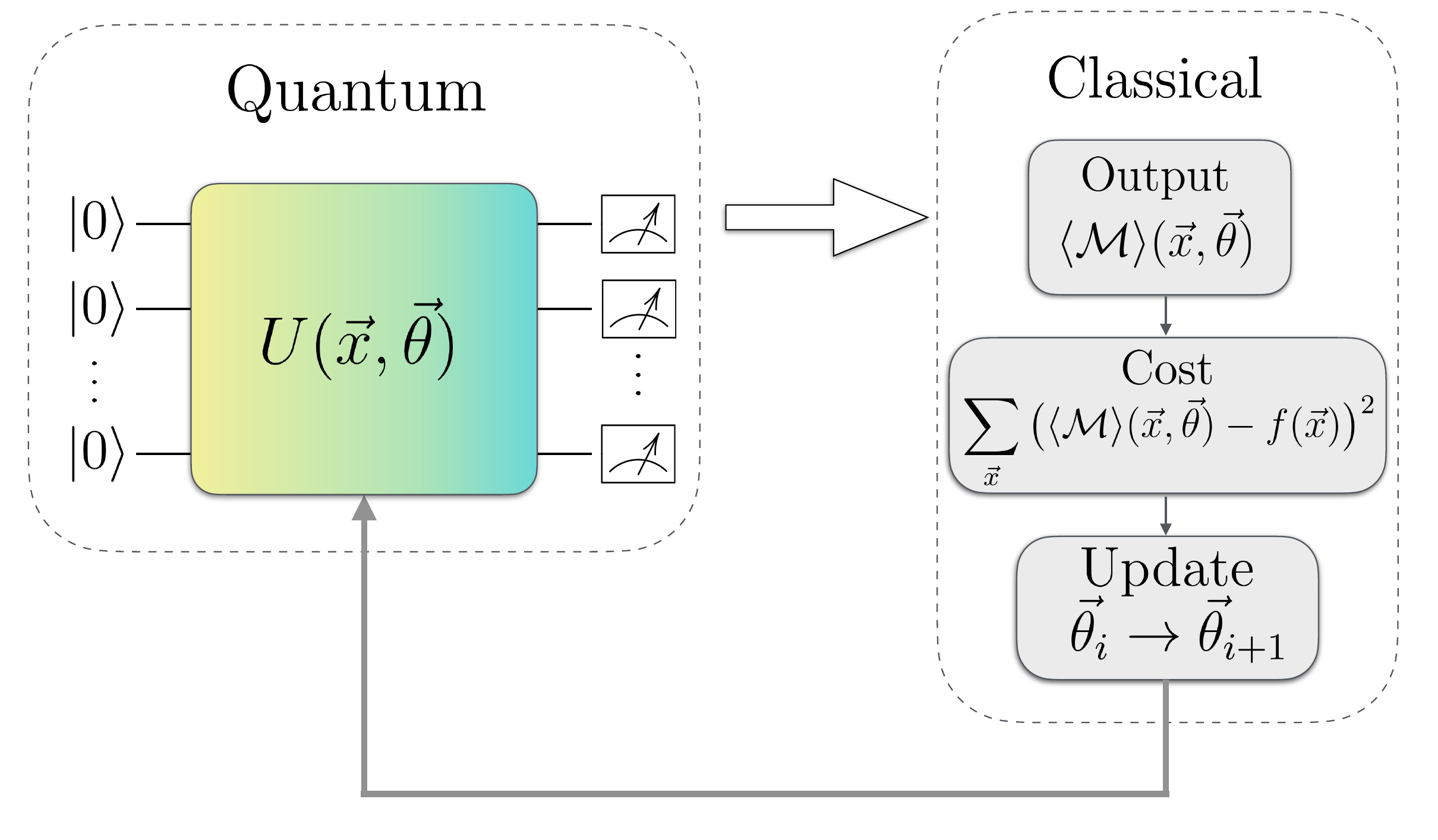}
\caption{Schematic representation of a quantum supervised learning model, which pertains to the family of Variational Quantum Algorithm (VQA). The picture divides the quantum and the classical part (dashed line boxes). The quantum part is composed of a quantum circuit with a given unitary $U$ that depends on the data point $\Vec{x}$ and the trainable parameters $\Vec{\theta}$. The classical part consists of building an expectation value of a given observable $\mathcal{M}$, introducing it into a cost function (in our case, tailored for the classification of a target function $f(\Vec{x})$) and optimize it with respect to the parameters via a classical subroutine.}
\label{fig:VQA}
\end{figure}

 This section reviews the quantum supervised learning model for fitting one-dimensional functions with the re-uploading strategy. We use a VQA fed with classical data, as we present in Fig. \ref{fig:VQA}. Using a training dataset $\{x\}$ with corresponding function images $f({x})$, we generate a quantum circuit for each data point by using encoding gates $S(x)$ and parameterized gates $A{(\vec{\theta}})$, which act as trainable gates. By measuring the expectation value of some operator $\mathcal{M}$, we obtain a Fourier series in the $\{x\}$ domain. After optimizing the parameters $\vec{\theta}$, we introduce test data points into the circuit to obtain a function prediction.

\begin{figure}[t]                         
\centering
\includegraphics[width=1\columnwidth, angle=0]{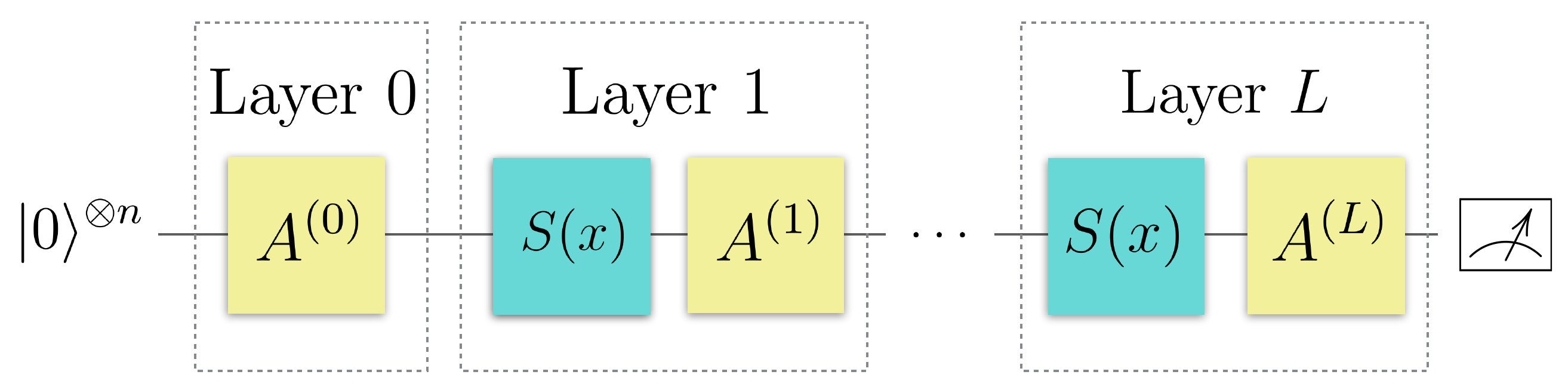}
\caption{General structure of a data re-uploading procedure. The 0th layer is used to generate an initial superposition. The other layers contain an encoding gate $S(x)$ and a processing gate $A(\Vec{\theta}_l)$, containing parameters which are optimized.}
\label{fig:DR}
\end{figure}

A widely used technique to define a PQC ansatz consists of defining a layer: a subcircuit composed of encoding and processing quantum gates. This structure is repeated $L$ times along the circuit~\cite{DR, perez2021one} (see Fig. \ref{fig:DR}). We define the general circuit layer $l$ with the data re-uploading encoding as
\begin{equation}
    U_{0} \equiv A(\Vec{\theta_0}), \quad
    U_{l}\equiv A(\vec{\theta_{l})}S(x),
    \label{eq:layer_def}
\end{equation}
where $A(\vec{\theta_{l})}$ and $A(\Vec{\theta_0})$ are general $n$ qudit unitary gates, with free parameters $\vec{\theta}$, that act as processing step.
For the encoding gate, a unitary gate $S(x)= e^{i x H}$ is used, where $H$ is an arbitrary encoding Hamiltonian, and $x$ is the data point considered. We assume that $H$ is diagonal since its single value decomposition is $S(x)=V^{\dagger}e^{ix\Sigma}V$, where $V$ is a unitary gate that can be re-absorbed by the processing gate $A({\vec{\theta_l)}}$, and $\Sigma$ is a diagonal matrix with the eigenvalues of $H$ (see App. \ref{app:Fourier} for the detailed derivation). The circuit layer can be interpreted as a quantum analogy of a neuron from a classical neural network because both structures are repeatedly fed with data. This is why these models are sometimes called quantum neural networks. The aforementioned quantum circuit ansatz has been used in prior works to prove that we obtain a one-dimensional truncated Fourier series as the output of these models~\cite{Effect_data_encoding,gil2020input}.

\begin{definition}
\textit{One-dimensional truncated Fourier series.}
    A truncated Fourier series is an expansion of a periodic real-valued function that can be expressed as a sum of sines and cosines. In the exponential form, it takes the following structure:
  \begin{equation}
      f(x) = \sum_{\omega = -D }^{D} c_{\omega} e^{ix\frac{\pi}{K}\omega},
  \end{equation}
  where $K$ is a half of the function period, $D = \text{max}(\Omega)$ is the Fourier series degree and $\omega \in \mathbb{N}$ are the multiples of the fundamental frequency $\frac{\pi}{T}$. The frequency spectrum is given by $\Omega = \{\omega \frac{\pi}{K}\}_{\omega}$. The coefficients $c_{\omega} \in \mathbb{C}$, fulfill the condition $c_{\omega} = c^{*}_{-\omega}$. A Fourier series with enough degree $D$ can approximate any continuous, square-integrable function. 
\end{definition}

By computing the expectation value of an observable $\mathcal{M}$ in the output state of the PQC, we obtain an approximation of $f(x)$. For simplicity, when we use more than one qudit in the PQC, we measure only one of them. Then, $\mathcal{M}= \mathbf{M}\otimes \mathbb{I} \otimes...\mathbb{I}$, where $\mathbf{M}$ is a single-qudit observable. We obtain (see App. \ref{app:Fourier} for the details)
\begin{equation}
    \langle \mathcal{M} (x)  \rangle = \sum_{\omega \in \Omega_{\textbf{kk'}}}
    c_{\omega} \ e^{i x \omega},
\end{equation}
with 
\begin{multline}
    c_{\omega} = \sum_{\substack{\mathbf{k,k'}=1\\ \Omega_{\textbf{kk'}} = \omega}}^{N} \left( \sum_{i=1}^{N} \mathcal{M}_{i} A_{ik_{L}}^{(L)} A_{ik'_{L}}^{*(L)}\right) \\ \times A_{k_{1}1}^{(0)}  A_{k'_{1}1}^{*(0)} \prod_{p=2}^{L}  A_{k_{p}k_{p-1}}^{(p-1)} A_{k'_{p}k'_{p-1}}^{*(p-1)},
    \label{eq:coef}
\end{multline}
where $N=d^{n}$ is the dimension of the Hilbert space of $n$ qudits with local dimension $d$, $L$ is the number of layers, and $\mathcal{M}_i$ are the eigenvalues of the observable, which we assume to be diagonal. We use the multi-index notation $\textbf{k} \equiv \{k_1, k_2,...k_L\} \in [N]^L$, with $k_i \in \{1,..N\}$. Also, we introduce the multi-index sum defined by $\Lambda_\textbf{k} \equiv (\lambda_{k_1}+\lambda_{k_2}+...\lambda_{k_L})$, where $\lambda_{k_i}$ is an eigenvalue of the encoding Hamiltonian $H$ in the layer $i$. With this, the frequency spectrum of the model is given by 
\begin{equation}
\begin{split}
    &\Omega_{\textbf{kk'}}\equiv  \{\Lambda_{\mathbf{k}}- \Lambda_{\mathbf{k'}}\} = \\ & \{\lambda_{k_{1}} + \lambda_{k_{2}} + \cdots +\lambda_{k_{L}} - \left(\lambda_{k'_{1}} + \lambda_{k'_{2}} + \cdots +\lambda_{k'_{L}}\right)\}.
\label{eq:freq}
\end{split}
\end{equation}
 The frequency spectrum is fully characterized by the eigenvalues of the encoding Hamiltonian, and the coefficients rely on the trainable gates and their parameters. Notice that all combinations of the multi-indices $\textbf{k}$ and $\textbf{k'}$ that generate a frequency $\Omega_{\textbf{kk'}}=\omega$ give us a different contribution to the coefficient $c_{\omega}$.

So far, this model accepts a general encoding Hamiltonian $H$, but let us take the following choice for practical purposes: using a Hamiltonian with the $N$-dimensional spin $S_{z}$ eigenvalues, which we name the ``spin-like" encoding. For one qubit we use $H=\frac{1}{2}\sigma_{z}$, with eigenvalues $\pm 1/2$; for one qutrit, the analogous Hamiltonian of a spin-1 Hamiltonian, which eigenvalues are $\pm 1, 0$, etc. We use a global $S_z$ Hamiltonian tailored to the dimension of the quantum circuit $N= d^n$. For example, for a system of two qubits ($N= 2^2 = 4$), $H$ is the spin-3/2 Hamiltonian, which acts on the whole circuit instead of using $H = \frac 12 \sigma_z \otimes \frac 12 \sigma_z$. This encoding will be the same used with a ququart, a qudit of $d=4$ (with the same value of $N$ as for two qubits). Using this particular encoding, the positive frequency spectrum, emerging from Eq. \eqref{eq:freq}, is $\Omega = \lbrace 0, \cdots, (N-1)L -1, (N-1)L \rbrace$, where $L$ is the number of layers in the circuit. The negative frequencies are also included in the spectrum, which is symmetric by its construction. For simplicity, from now on, we drop the label for multi-indices in the frequency spectrum: $\Omega \equiv \Omega_{\textbf{kk'}}$. The degree of the Fourier series, with this particular encoding, is given by $D = (N-1)L$.

The spectrum generated with the ``spin-like" encoding only contains integer frequencies. For instance, if the function we are fitting requires semi-integer or real frequencies a proper approximation cannot be achieved, regardless of the number of layers used. We can tackle this issue by introducing a re-scaling parameter $\eta$ into the encoding gate: $S(x)=e^{i x \eta H}$. This parameter is optimized together with the rest of the free parameters of the circuit, as we explain in App. \ref{app:rescaling}. With this re-scaling factor, the whole Fourier series spectrum is multiplied by $\eta$, and the degree of the Fourier series becomes $D=\eta(N-1)L$. We can introduce more fine-tuning in the frequency spectrum by using a different $\eta_{i}$ for each encoding gate in the circuit layers. This extension is closer to the original idea from the data re-uploading work \cite{DR} but should be treated carefully, because methods with re-scaling parameters may lead to good expressibility but overfitting and poor generalization bounds for more complex tasks \cite{6795928}. Nevertheless, when applied appropriately, the re-scaling factor can serve as an effective hyperparameter for promoting generalization in quantum kernel models \cite{canatar2022bandwidth}. It is important to note that the encoding strategy plays a significant role in determining the characteristics of the frequency spectrum. For instance, in Ref. \cite{Shin2022ExponentialDE}, the authors propose a technique that utilizes re-scaling factors to generate exponentially more frequencies.

\section{Multidimensional Fourier series with quantum circuits}\label{sec:Multi FS}

In this section, we present different ansatzes that generate multidimensional Fourier series and explore the scaling of the QML models' performance with the dimensions of the input data. To do it, we study the expressibility of the multidimensional Fourier model, meaning the type of functions that the model can generate. Hence, we expand the function fitting formalism to multidimensional data, opening the possibility to explore more complex problems with these quantum models.

\begin{definition}
\textit{Multidimensional truncated Fourier series.} 
    The generalization of a one-dimensional truncated Fourier series to $M$-dimensional data is given by 
\begin{equation}
    f(\Vec{x}) = \sum _{\omega_1,\omega_2,...,\omega_M = -D}^{D} c_{\Vec{\omega}} e^{i \Vec{x}\cdot \Vec{\omega}},
\end{equation}
where $D = \text{max}(\omega_1,\omega_2,...,\omega_M)$ is the degree of the Fourier series. The data $\Vec{x} = (x_1,x_2,...,x_M)\in \mathbb{R}^M$ and the frequencies $\Vec{\omega} = (\omega_1, \omega_2,..., \omega_M)\in \mathbb{Z}^M$ are represented by $M-$dimensional vectors, and $\Vec{x}\cdot \Vec{\omega}$ is the scalar product. The coefficients  $c_{\omega_1, \omega_1,..., \omega_M} \in \mathbb{C}$ fulfill the relation $c_{\omega_1, \omega_1,..., \omega_M} = c_{-\omega_1, -\omega_1,..., -\omega_M}^*$.  
\end{definition}

The $M$-dimensional Fourier series contains substantially more coefficients than a series with only one dimension. For a given degree $D \in \mathbb{N}$, the number of independent coefficients $c_{\Vec{\omega}}$ is
\begin{equation}
    N_c = \frac{(2D+1)^M -1}{2}+1.
    \label{eq:num_coef}
\end{equation}
As an example, for degree $D=1$ and data-dimension $M=2$, according to Eq. \eqref{eq:num_coef}, we have $5$ coefficients: $c_{00}, c_{01}, c_{10}, c_{11}, c_{-11}$, while the other ones are constrained by $c_{\omega_1 \omega_2} = c_{-\omega_1 -\omega_2}^*$. The degrees of freedom $\nu$ (abbreviated as DOF) of the Fourier series are the number of independent variables needed to fully characterize a set of coefficients of a series with a given degree $D$. They account for the real and imaginary parts of each coefficient, except the one associated with the zero frequency, which only contains a real part. Therefore, the DOF of an $M$-dimensional Fourier series are given by
\begin{equation}
    \nu \equiv 2N_c-1 = \big(2D+1)^M.
    \label{eq:degrees_freedom}
\end{equation}
Depending on the circuit ansatz, the number of layers required to achieve enough freedom to represent an arbitrary series will vary. In the following subsection, we propose four strategies: the Line, Parallel, Mixed, and  Super-parallel Ansatzes. We assume that the processing gates are general unitary transformations in all models. Hence, the number of parameters they contain is  $N^2-1$, determined by the dimension of SU$(N)$, with $N=d^n$ being the dimension of the quantum circuit. We acknowledge that this approach may not be practical in terms of trainability. However, we utilize this strategy to explore the limits of the models.

\subsection{Line ansatz}

\begin{figure*}[t!]                        
\centering
{\includegraphics[width=1\columnwidth, angle=0]{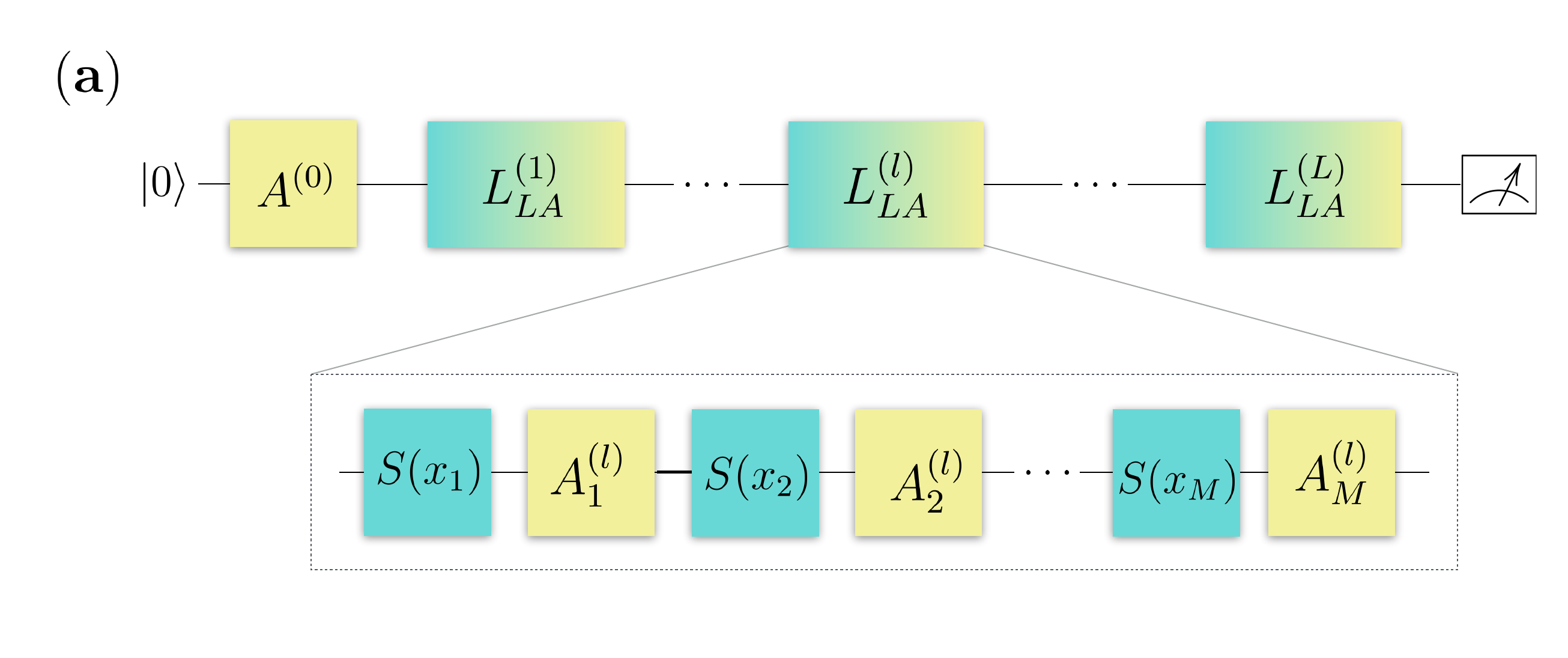}} 
{\includegraphics[width=1\columnwidth, angle=0]{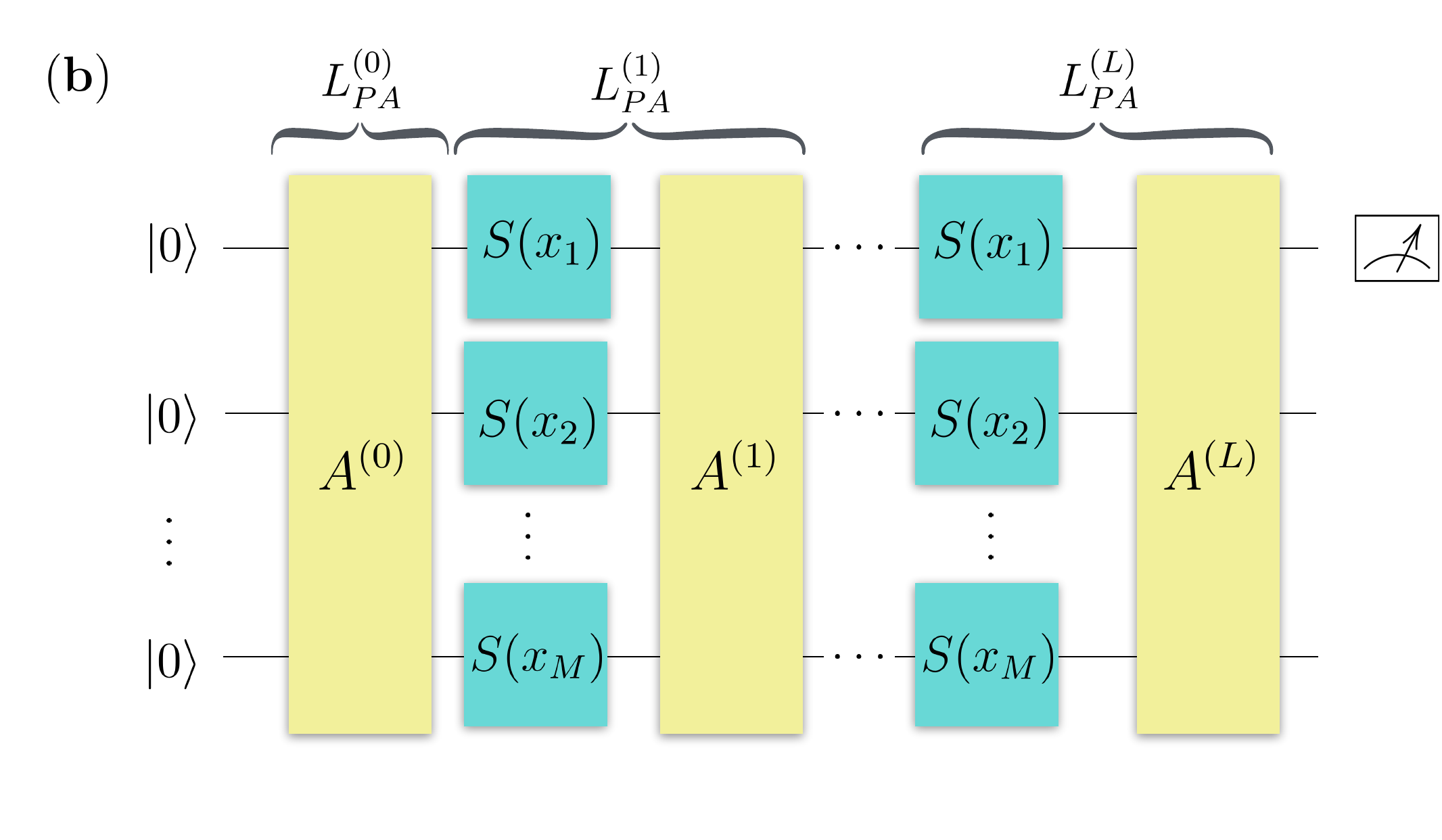}}\\
{\includegraphics[width=1\columnwidth, angle=0]{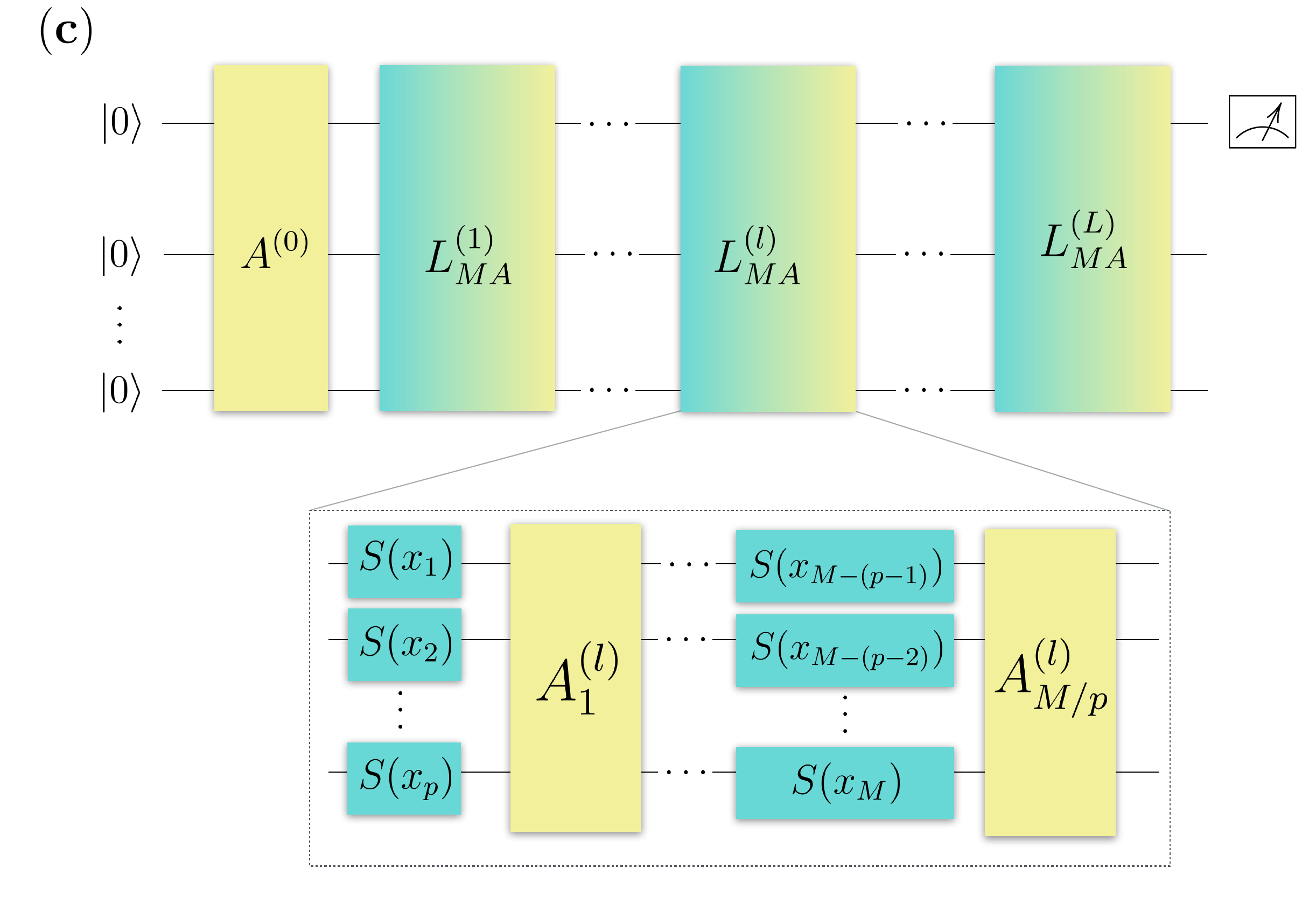} }
{\includegraphics[width=1\columnwidth, angle=0]{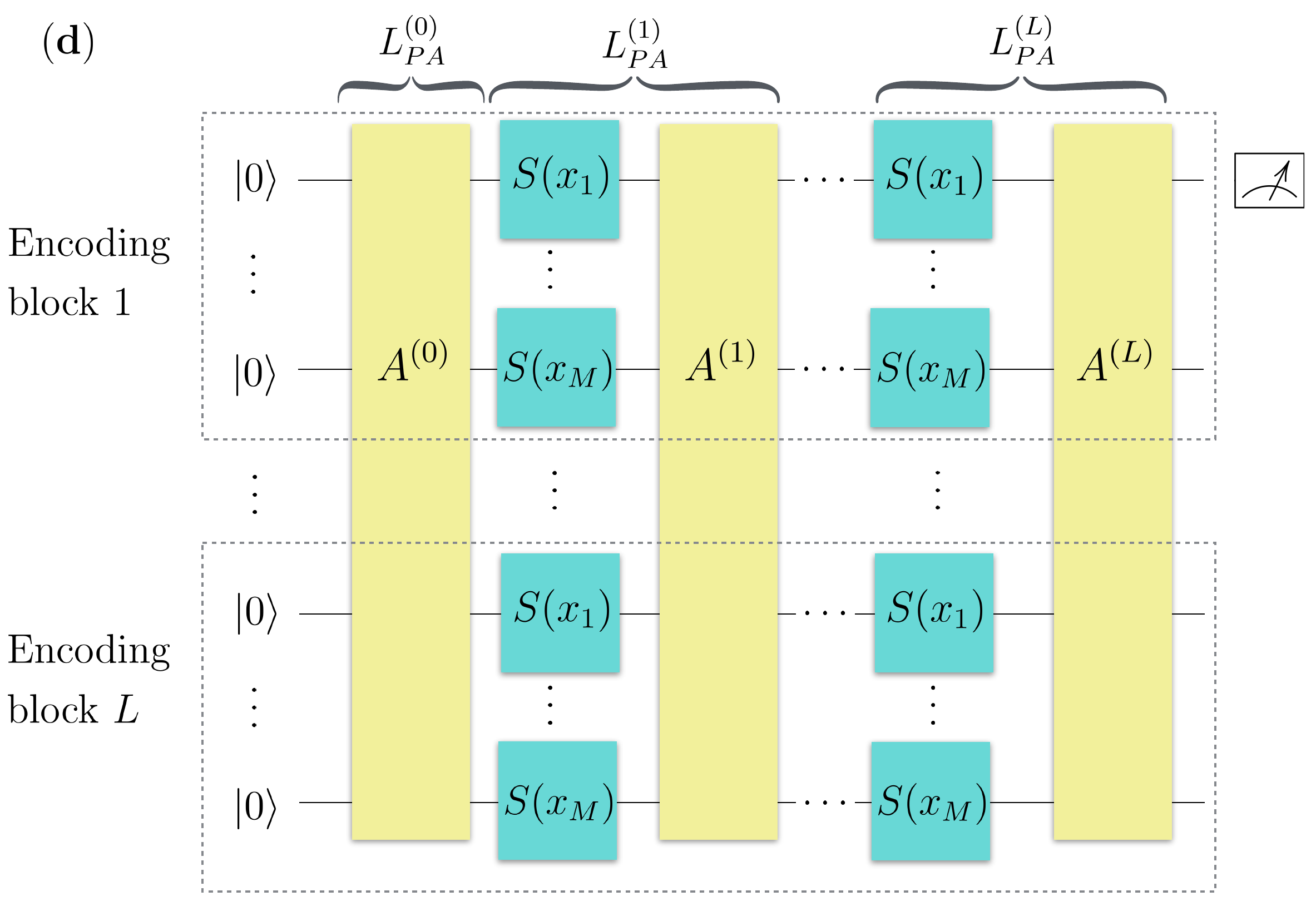}}
\caption{Quantum circuit ansatzes of the models. The Line ansatz (\textbf{a}) encodes each data feature in a single qudit. Thus, the circuit depth grows linearly with the total number of features $M$. The Parallel ansatz (\textbf{b}) encodes the $M$ features in $M$ qudits instead. The Mixed ansatz (\textbf{c}) combines the two approaches by distributing the features encoding between $p<M$ qudits and uses more gates to introduce several features in each layer. Finally, the Super-Parallel ansatz (\textbf{d}) uses $L$ layers and $L$ encoding blocks per qudit, therefore requiring $d= ML$ qudits.}
\label{fig:ansatzes}
\end{figure*}

The Line Ansatz (LA) encodes all data dimensions in a single qudit. The structure of the model is shown in Fig. \ref{fig:ansatzes} (\textbf{a}). Each layer $L^{(l)}$ of the LA encodes $M$ data features as  
\begin{equation}
    L^{(l)}(\vec{x},\vec{\theta})_{\text{LA}}\equiv \prod_{m=1}^{M}S(x_{m}) A_{m}^{(l)}(\Vec{\theta}_{l,m}),
\end{equation}
where $M$ is the dimension of the dataset, $x_{m}$ the data features and $\Vec{\theta}_{l,m}$ are the processing parameters corresponding to feature $m$ from layer $l$. After the encoding step, the processing gate is applied to avoid the collapse of the data in a single variable (see App. \ref{app:LA} for details). 

Another strategy to encode multidimensional data in one qudit is to use non-commuting gates, meaning $[S_1(x), S_2(x) ] \neq 0$. For example, in the two-dimensional qubit case, we could use $S_1(x_1) = R_y(x_1)$ and $S_2(x_2)=R_z(x_2)$. Given that $S_1$ and $S_2$ do not commute, the single value decomposition of the encoding gate is given by
\begin{equation}
    S_1(x_1) S_2(x_2) = V^{\dagger}_1 \Sigma(x_1) V_1 V^{\dagger}_2 \Sigma(x_2) V_2.
\end{equation}
This can be interpreted as adding an extra layer $V_1V_2^{\dagger}$ in between the two encoding gates that does not contain trainable parameters and also avoids the model from interpreting the data as one-dimensional.

The number of parameters to be optimized in the LA ansatz is given by
\begin{equation}
    N_p^{(LA)}= (ML+1)(N^2-1) \sim MLN^2.
\end{equation}
As mentioned above, $N=d$ is the dimension of the circuit and the dimension of the encoding gate $S(x)$. 
Thus, the parameters grow linearly with the number of layers $L$ and the data dimension $M$, and quadratically with the circuit dimension $N$. 

\begin{figure*}
\centering
\includegraphics[width=2\columnwidth]{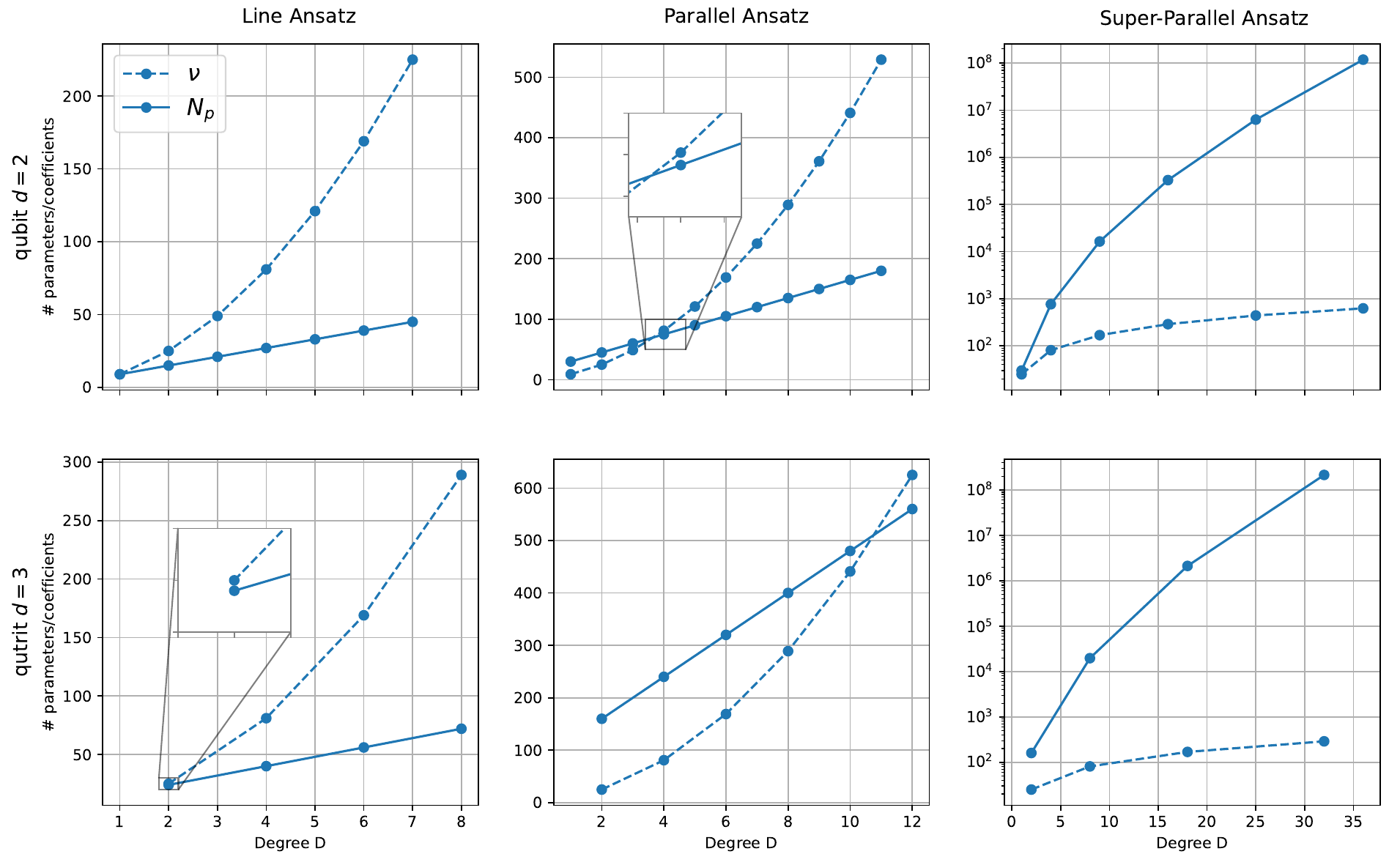}
    \caption{Comparison of the degrees of freedom condition for a two-dimensional Fourier series for the Line (first column), Parallel (second column), and Super-parallel (third column) ansatzes using qubits (first row) and qutrits (second row). The dashed lines represent the degrees of freedom $\nu$ of the Fourier series produced by the models plotted against its degree $D$, while the solid line depicts the number of trainable parameters $N_p$ in the circuit ansatzes that generate a Fourier series of degree $D$ with $L$ layers. The Parallel Ansatz fulfils the degrees of freedom condition for higher-degree Fourier series compared to the Line Ansatz, being the gap wider when using qutrits. In particular, the condition for the Parallel Ansatz is fulfilled until $D=3$ and $D=10$ for qubits and qutrits, respectively. In the Super-Parallel ansatz, plotted on a logarithmic y-axis scale, we observe a faster increase in the number of parameters compared to the degrees of freedom. }
    \label{fig:comparisson}
\end{figure*}

\subsection{Parallel ansatz}

The parallel ansatz (PA) encodes each data feature in a different qudit with a single-qudit gate, therefore, we require $n=M$ qudits to encode all $M$ data features. The encoding is followed by a processing $M-$qudit gate, as shown in Fig. \ref{fig:ansatzes} (\textbf{b}). 
We define each layer as
\begin{equation}
     L^{(l)}(\vec{x},\vec{\theta})_{\text{PA}}\equiv \left(\bigotimes_{m=1}^{M}S(x_{m})\right) A^{(l)}(\Vec{\theta_{l}}),
\end{equation}
where $A^{(l)}(\Vec{\theta_{l}})$ is a $N\times N$ general processing unitary, with $N= d^M$. This ansatz contains a total number of parameters
\begin{equation}
    N_p^{(PA)}= (d^{2M}-1)(L+1) \sim d^{2M}L,
\end{equation}
which grows exponentially with the number of features $M$. 

If single-qudit gates are used in the processing steps instead of general multi-qudit gates, then a product of $M$ one-dimensional Fourier series is generated. Therefore, entanglement must be included in the processing gates to obtain a genuine multidimensional series (see App. \ref{app:PA} for more details).

\subsection{Mixed ansatz}

Now we consider a mix between the two ansatzes previously discussed, named the Mixed Ansatz (MA). It divides the data features into different batches and uses $p$ qudits for different sets (see Fig. \ref{fig:ansatzes} (\textbf{c})). Taking $p\leq M$ qudits, the data features are distributed in $p$ qudits, as in the PA, and more encoding layers can be used for each qudit if required, as in the LA. For $p=1$, MA behaves like LA, and for $p=M$ it behaves like PA.  More formally, we define a layer of the MA as
\begin{equation}
     L^{(l)}(\vec{x},\vec{\theta})_{\text{MA}}\equiv \prod_{k=1}^{\lceil M/p\rceil}\left(\bigotimes_{m=1}^{p}S(x_{m})\right) A_{k}^{(l)}(\Vec{\theta_{l,k}}),
\end{equation}
and the number of parameters in this model is given by
\begin{equation}
    N_p^{(MA)} = (d^{2p}-1) (\lceil M/p \rceil L+1).
\end{equation}
If $M/p$ is not an integer, we use fewer encoding gates on the layers' last encoding block.

\begin{figure*}[t!]                        
\centering
\includegraphics[width=1.5\columnwidth, angle=0]{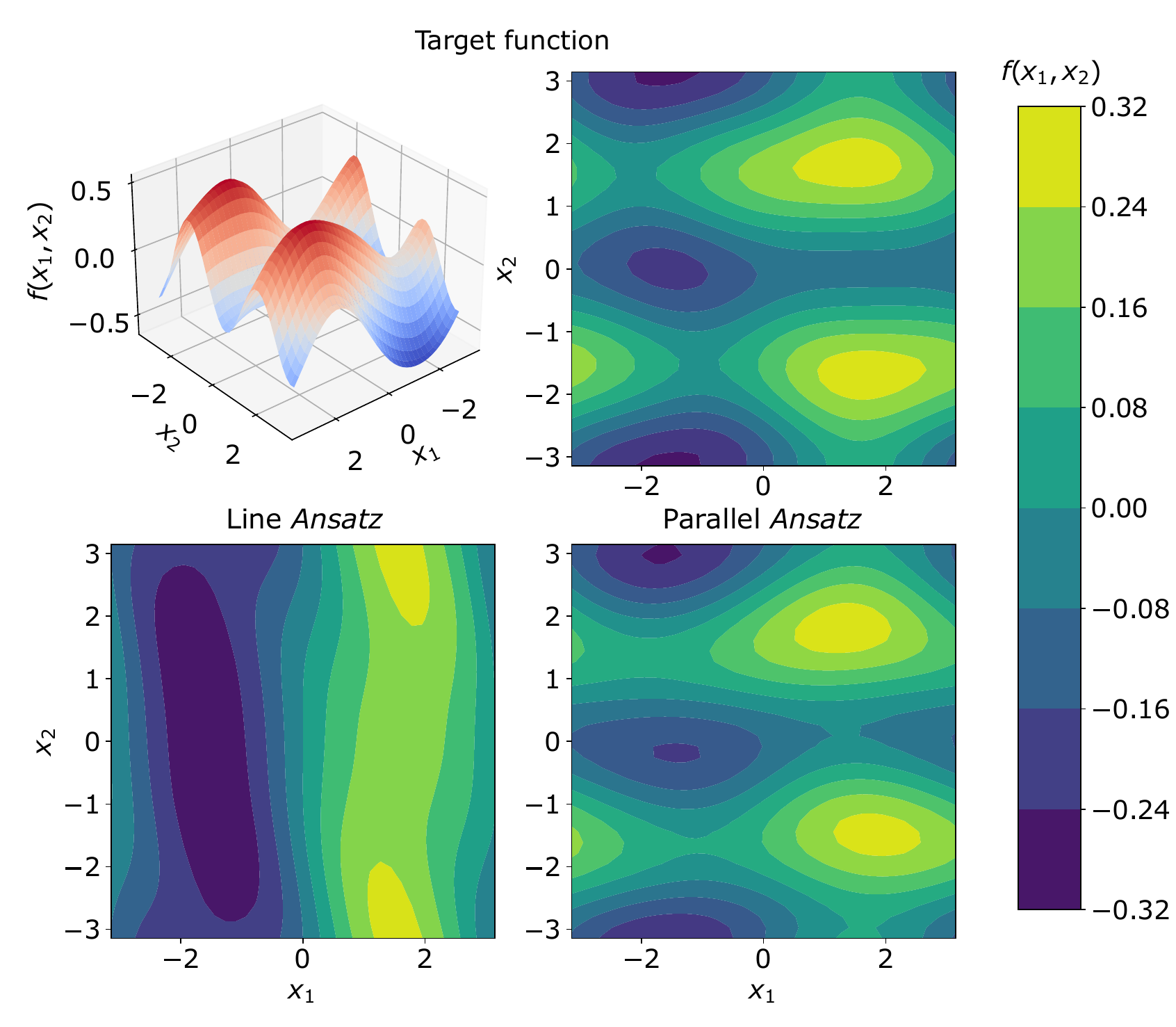}
\caption{Simulation the Line and Parallel ansatzes for fitting a Fourier series of degree $D = 2$. The target function in the trigonometric form is given by $f(x_1,x_2) = -0.02+ 0.04\cos(2x_1+x_2)+ 0.25\sin x_1-0.3\cos 2x_2-0.1\sin(x_1-x_2)$. We use the Nelder-Mead method as a classical optimization subroutine with $500$ training and $1500$ testing data points.  In the Line ansatz, we use one qubit and $L=2$. The accuracy obtained is $38.53\%$ and, as we can see, the model does not capture the structure of the target function. In the  Parallel ansatz, we use two qubits and $L=2$. The accuracy obtained is  $95.63\%$, and the predicted and target functions have a similar structure. In this regime, the Line ansatz model does not fulfil the DOF condition, while the Parallel does. This is reflected in the accuracy of the simulations.}
\label{fig:fitting_2d}
\end{figure*} 

\subsection{Super-Parallel ansatz}
Finally, we provide an ansatz that includes layers in depth and width directions of the quantum circuit. In other words, it has the same layer structure of re-uploading models but with $L$ encoding blocks per layer. Each block contains $M$ single-qubit encoding gates (one for each data feature). As shown in Fig. \ref{fig:ansatzes} (\textbf{d}), a total of $n = LM$ qudits are required for this ansatz. Each layer is given by 
\begin{equation}
    L^{l}(\vec x, \vec \theta )_{SP} \equiv \left(  \bigotimes_{i=1}^{L}  \left( \bigotimes_{m=1}^{M}S(x_{m})\right) \right) A^{(l)}(\Vec{\theta_{l}}),
\end{equation}
and the number of parameters in this model is
\begin{equation}
    N_p^{(SP)} = (L+1)(d^{2ML}-1)\sim Ld^{2ML}, 
    \label{eq:NP_SP}
\end{equation}
which grows exponentially with the input data dimension and the number of layers.

\section{Results and discussion}\label{sec:results}
To compare the models, we assume that all use singe-qudit encoding gates, although their extension to multi-qudit gates is straightforward (equivalent to finding an information unit of dimension $\Bar{d}=d^n$). The degree of the Fourier series generated by the first three ansatzes is determined by
\begin{equation}
    D = (d-1)L,
    \label{eq:degree_ansatzes}
\end{equation}
where $d$ is the dimension of the qudit(s) used in the model. The super parallel model outputs a Fourier series of degree
\begin{equation}
    D^{(SP)} = (d-1)L^2,
    \label{eq:degree_SP}
\end{equation}
which grows quadratic with the number of layers due to the use of multiple encoding blocks per layer. For detailed derivation and further information on the models, refer to App. \ref{app:LA}, \ref{app:PA}, \ref{app:MA}, and \ref{app:SP}. By specifying the qudit dimension $d$ and the number of layers $L$, we determine the Fourier series frequency spectrum $\Omega$ and its degree $D$. This allows us to calculate the degrees of freedom $\nu$ of the output series by plugging Eq. \eqref{eq:degree_ansatzes} and \eqref{eq:degree_SP} in Eq. \eqref{eq:degrees_freedom}. Our next step is to investigate whether we can fully represent the Fourier series coefficients by comparing the degrees of freedom $\nu$ with the number of independent parameters $N_p$ in the quantum circuit. The ansatzes are required to contain, at least, $\nu$ free parameters to generate any series coefficient, resulting in a condition  $N_p \geq \nu$. When this condition is not accomplished, the model is not general enough to approximate all possible Fourier series. However, having $\nu>N_p$ does not necessarily guarantee that we can fit a general series since the coefficient equations are highly coupled and non-linear (see Eq. \eqref{eq:coef}), although one might expect that, in general, $\nu>N_p$ is enough to approximate a given series. Our goal is to establish a lower bound in the worst-case scenario in terms of the circuit ansatz requirements to fit a general multidimensional Fourier series. 

The one-dimensional feature model, introduced in section \ref{sec:FS with quantum circuits}, possesses $\nu = 2(D+1) = 2(L(d-1)+1)$ degrees of freedom, and $N_p = (L+1)(d^2-1)$ free parameters. Both these quantities increase linearly with the number of layers and $N_p > N_c$ holds for all values of $L$ and $d$, which implies that the degrees of freedom requirements are satisfied in all cases.

For the line, parallel and mixed ansatz with a fixed qudit dimension, the degrees of freedom grow polynomially with the number of layers, $\nu \sim  L^M$, while the number of parameters grows proportionally to the layers used $N_p \sim L$, regardless of the data dimension. Therefore, at some point, the DOF will exceed the number of free parameters of the model.

In Fig. \ref{fig:comparisson} we represent the DOF condition for the Line, Parallel and Super-parallel ansatzes for a case of two-dimensional data with qubits and qutrits models. In the first place, we see that the Line Ansatz only fulfils the condition for the qubit model with $D=1$.  Taking into account the asymptotics in the degrees of freedom condition $N_p \geq \nu $ and assuming that $\nu \sim (2DL)^M$ and $N_{p}^{(LA)}\sim MLd^2$, we obtain the following condition:
\begin{equation}
   d \lesssim \big( \frac{M}{2^ML^{M-1}}\big)^{1/ (M-2)} \xrightarrow[M \to \infty]{} \frac{1}{2L}.
\end{equation}
This suggests that simply increasing the qudit dimension is insufficient to achieve the desired number of parameters. In particular, for large data dimensions, the condition can not be fulfilled because $d\leq 1/2L$ is an impossible condition for $L>1$. Therefore, one-qudit models have limited power for fitting multidimensional functions.

For the Parallel ansatz depicted in the second row of Fig. \ref{fig:comparisson}, we see that by increasing the qudit dimension, we arrive at higher Fourier series Degrees. Indeed, by running the asymptotics in the DOF condition considering $N_p^{(PA)}\sim Ld^{2M}$, we achieve
\begin{equation}
    d\gtrsim 2L^{\frac{M-1}{M}}\xrightarrow[M \to \infty]{}2L,
\end{equation}
which implies that in order to satisfy the inequality, we would need to use qudit dimensions that increase with the number of layers. In particular, in the limit of large dimensional datasets, $d$ has to grow proportional to $L$. This indicates that the model remains universal, in the sense that can generate any arbitrary Fourier series, as long as $d$ can grow with $L$. In that case, the complexity of the problem would shift to finding systems with arbitrarily large qudit dimensions to fit generic Fourier series, which can be resource-demanding for some technologies.

Fig. \ref{fig:fitting_2d} depicts the results of simulation for models fitting a two-dimensional Fourier series of degree $D=2$ with the Line and Parallel ansatzes. Since for two-dimensional data, the LA does not accomplish the DOF condition, the simulation does not find suitable parameters to fit the coefficients. On the contrary, the parallel model approximates with high accuracy the target function with only two layers, because meets the DOF requirements. For more details on two and three-dimensional models see Ref. \cite{Gratsea2022T}, where expressibility is analyzed in detail.

The limitation that we exhibit for the Line and Parallel ansatzes is that we do not have enough free parametrization in the Hilbert space of the PQC to accomplish the DOF that multidimensional Fourier series require. Therefore, we can only fit functions up to a certain degree. One might assume that the problem could be solved by adding more layers, which introduces more trainable parameters in the PQC to match the DOF required for the desired set of coefficients. However, the addition of more layers also increases the output Fourier series' degree $D$ (see Eq. \eqref{eq:degree_ansatzes}), which requires more coefficients and thus more degrees of freedom. Therefore, the problem cannot be resolved by merely increasing the number of layers because this also raises the required degrees of freedom of the output model.

The Super-parallel ansatz has a number of qudits that grows with the number of layers and the data dimension. In this scenario, the degrees of freedom scale as $\nu \sim (2dL^2)^M$. Substituting this into the inequality $N_p\geq \nu$ with the parameter count given in Eq. \eqref{eq:NP_SP} yields
\begin{equation}
d\geq 2^{\frac{1}{2L-1}}L^{\frac{2}{2L-1}}.
\end{equation}
It can be easily seen that this inequality is always satisfied. An example of this can be observed in the third row of Fig. \ref{fig:comparisson}, where we can notice that the number of parameters grows more rapidly than the degrees of freedom of the resulting model function, making the model capable of fitting any arbitrary Fourier series.  There is an increasing gap between the number of parameters and degrees of freedom, which opens the possibility of finding more sophisticated Super-parallel encodings that can employ lesser variables while remaining universal, for instance, by using non-general unitary gates or fewer qudits.

\section{Conclusions}\label{sec:conclusions}

In this work, we have explored how to generate multidimensional Fourier series with parameterized quantum circuits. These series emerge naturally from the expectation values of quantum operators with a particular data encoding. We have compared the degrees of freedom of a general Fourier series of a given degree to the number of free parameters in the circuit, which we refer to as the DOF condition and provides insight into the model expressibility. We provide a trade-off between the number of qudits, circuit depth (measured with the number of layers of the circuit), data dimension, and local qudit dimension.

Current quantum computers can use higher energy states to perform high-dimensional quantum computation. Apart from this, the use of a general formalism for qudits is motivated by the possibility of exploring larger Hilbert spaces (with more extensive parametrization), which gives more freedom when fitting the desired set of coefficients.

 For one-dimensional data, the DOF requirement is always accomplished. However, for higher-dimensional data, the degrees of freedom grow exponentially with the data dimensions, which can be problematic for some models that may not be able to keep up with this rapid growth. For example, single-qudit models have limited power in the expressivity of multidimensional data, because they lack the appropriate parametrization. However, multi-qudit models can approximate functions up to a higher degree, which can be used as an appropriate approximation for some problems. 

We can always find a model that satisfies the DOF condition by using large qudit dimensions or a substantial number of qudits, then being fully expressive. Therefore, multidimensional quantum learning models can be considered universal, as with sufficient parametrization, they can fit any arbitrary Fourier series.  Potential issues may appear with trainability and generalization resulting from using such a large parametrization. However, we speculate that for most problems, such an extensive parametrization would not be necessary nor practical in terms of trainability.

The Line, Parallel and Mixed ansatzes exhibit an inductive bias towards limited-band functions, which contain low frequencies. Further work needs to be done for studying how the inductive bias of quantum learning models \cite{kubler2021inductive, peters2022generalization} varies with the PQC used. It would be interesting to benchmark the performance, trainability, and generalization capabilities of the ansatzes against a classical surrogate model \cite{Schreiber2022C, Landman2022C}. Another open question is how to determine the level of redundancy necessary in the output Fourier series degree when the target function is not known beforehand. 

We aim for our work to contribute to the understanding of QML with classical multidimensional data and the further exploration of more sophisticated embedding strategies. \\

\section*{Code and data availability}

All code used in this work can be found on the GitHub repository \url{https://github.com/bsc-quantic/fourier}.

\section*{Acknowledgements}\label{sec:acknowledgements}

This work has been financially supported by the Ministry of Economic Affairs and Digital Transformation of the Spanish Government through the QUANTUM ENIA project call – Quantum Spain project, and by the European Union through the Recovery, Transformation and Resilience Plan – NextGenerationEU within the framework of the Digital Spain 2026 Agenda. Also, we acknowledge the BSC's Quantic group members for the suggestions and discussions provided.

%%%%%%%%%%%%%%%%%%%%%%
% BIBLIOGRAPHY
%%%%%%%%%%%%%%%%%%%%%%%%%
\bibliography{biblio.bib}

\newpage

\onecolumngrid

\appendix
    \section{Data re-uploading for one-dimensional Fourier series} \label{app:Fourier}

This section explains how we generate one-dimensional Fourier series with a circuit of $n$ qudits of dimension $d$. For processing gate a general unitary is taken, of total dimension $N=d^n$ ($n$ qudits of dimension $d$):
\begin{equation}
    A^{(l)}=\left(\begin{array}{ccc}
        A_{11}^{(l)} & \cdots & A_{1N}^{(l)}  \\
        \vdots &  \ddots & \vdots \\
        A_{N1}^{(l)} & \cdots & A_{NN}^{(l)}
    \end{array} \right).
    \label{eq:gates}
\end{equation}
In contrast to the processing, the encoding gates are the same in all layers. Let's assume the following single-qudit encoding gate:
\begin{equation}
    S(x)=e^{i x H},
\end{equation}
where $H$ is a $N\times N$ Hermitian operator and $x$ is a one-dimensional data point from the function we want to represent. We assume that the encoding Hamiltonian $H$ is diagonal because, when taking the Singular Value Decomposition, $H=V^\dagger \Sigma V$, where $V$ and $V^{\dagger}$ are unitary matrices and $\Sigma$ is a diagonal matrix formed with the eigenvalues of $H$, the encoding gate becomes
\begin{equation}
    S(x) = e^{-ixV\Sigma V^{\dagger}} = \sum_{m=0}^{\infty} \frac{1}{m!} (-ix V\Sigma V^{\dagger})^m = 1+  \sum_{m=1}^{\infty} \frac{1}{m!}V (-ix\Sigma)^m V^{\dagger} = Ve^{-ix\Sigma }V^{\dagger} =VR(x)V^{\dagger} .
    \label{eq:absorb_unitaries}
\end{equation}
where we take into consideration that $V^{\dagger}V = \mathbb{I}$ and $\Sigma = \text{diag}(\lambda_1,...,\lambda_N)$.  The resulting diagonal encoding gate is
\begin{equation}
    R(x) =\text{diag}\left(e^{ix\lambda_{1}},\cdots,e^{ix\lambda_{N}}\right).
    \label{eq:encoding}
\end{equation}
Therefore, we assume, without loss of generality, a diagonal encoding matrix, since $V$ and $V^{\dagger}$ gates are re-absorbed in the definition of the general processing gates $A^{(l)}$. With this assumption, each layer $l$ is composed of the product
\begin{equation}
    L^{(l)}=A^{(l)}R(x) \qquad  L^{(0)}= A^{(0)},
    \label{eq:layer}
\end{equation}
which matrix elements are
\begin{equation}
    L_{ij}^{(l)}=\sum_{k=1}^{N}A_{ik}^{(l)}R_{kj} = A_{ij}^{(l)}e^{i x \lambda_{j}}, \qquad L_{ij}^{(0)}=A_{ij}^{(0)}.
    \label{eq:layer_elem}
\end{equation}
In general, for $L$ layers, the unitary transformation of the whole circuit is expressed as
\begin{equation}
    U_{ij}=  \sum_{k_{1},\cdots,k_{L}=1}^{N}   A^{(L)}_{i k_L}e^{-ix\lambda _{k_L}} A^{(L-1)}_{ k_L k_{L-1}}...A^{(1)}_{k_2k_1} e^{-ix\lambda_{k_1}} A^{(0)}_{k_1 k_0}.
\end{equation}
The initial state of the circuit is the zero state of dimension $N$. Therefore $|0\rangle = (1,0,\cdots,0)^T$. Thus, the state generated by the circuit becomes:
\begin{eqnarray}
    |\psi\rangle &=&  U|0\rangle^{\otimes n} ,\\
    \psi_{i} &=& U_{ij} \delta_{j1} = U_{i1}.
\end{eqnarray}
Putting it all together, we obtain
\begin{equation}
    \begin{split}
       \psi_i = 
   \sum_{k_{1},\cdots,k_{L}=1}^{N}   e^{-ix(\lambda _{k_1}+...+\lambda _{k_L})} A^{(L)}_{i k_L} A^{(L-1)}_{ k_L k_{L-1}}...A^{(1)}_{k_2k_1} A^{(0)}_{k_1 1},
    \end{split}
\end{equation}
By introducing the multi-index notation $ \textbf{k} \equiv \{ k_1,...,k_L\} \in [N]^L$ and the multi-index sum $\Lambda_\textbf{k} = \lambda _{k_1}+...+\lambda _{k_L}$,we re-express the $i$-vector state: 
\begin{equation}
     \psi_i =\sum_{\textbf{k}\in [N]^L}  e^{-ix\Lambda_\textbf{k}} A^{(L)}_{i k_L}  A^{(L-1)}_{ k_L k_{L-1}}...A^{(1)}_{k_2k_1} A^{(0)}_{k_1 1}. 
\end{equation}
Each multi-index $\textbf{k}$ is a possible combination of $L$ indices and each run from $1$ to $N$. The multi-index sum $\Lambda_\textbf{k} = \lambda _{k_1}+...+\lambda _{k_L}$ is a sum that has $|\textbf{k}| = N^L$ possible values, given by all the possible combinations of $\lambda_{k_{i}}$. 

Now we compute the expectation value of a given observable $\mathcal{M}$ in this state. The observable can be diagonal by the same argument used for the encoding gate $S(x)$, explained above. The eigenvalues of any observable are real, so $\mathcal{M}_{ii}^{*}=\mathcal{M}_{ii} \equiv \mathcal{M}_{i}$. All together,
\begin{equation}
    \langle \mathcal{M} \rangle = \langle \psi | \mathcal{M} |\psi\rangle = \sum_{i=1}^{N}U_{i1}^{*} M_{i} U_{i1} = \sum_{\substack{k_{1},\cdots,k_{L}=1 \\ k'_{1},\cdots,k'_{L}=1 }}^{N}
    e^{i x \left(\Lambda_{\mathbf{k}}- \Lambda_{\mathbf{k'}}\right)} \
    \left( \sum_{i=1}^{N} M_{i} A_{ik_{L}}^{(L)} A_{ik'_{L}}^{*(L)}\right) A_{k_{1}1}^{(0)}  A_{k'_{1}1}^{*(0)} \prod_{p=2}^{L}  A_{k_{p}k_{p-1}}^{(p-1)} A_{k'_{p}k'_{p-1}}^{*(p-1)}.
\end{equation}

For each set of $\mathbf{k}$ and $\mathbf{k'}$ parameters, we generate a particular frequency $\Omega_{kk'} = \Lambda_{\mathbf{k}}- \Lambda_{\mathbf{k'}}$. Notice that different $\mathbf{k,k'}$ choices can give the same frequency. Also, to obtain the opposite sign value $-\Omega_{kk'}$, one needs to exchange the $\mathbf{k}$ and $\mathbf{k'}$ indices. Therefore, we can group the coefficients that generate the same frequencies from the above expression:
\begin{multline}
    \langle \mathcal{M} \rangle = \sum_{\omega\geq 0}  \sum_{\substack{\mathbf{k,k'}=1\\ \Omega_{kk'} = \omega}}^{N}
    e^{i x \omega} \
    \left( \sum_{i=1}^{N} M_{i} A_{ik_{L}}^{(L)} A_{ik'_{L}}^{*(L)}\right) A_{k_{1}1}^{(0)}  A_{k'_{1}1}^{*(0)} \prod_{p=2}^{L}  A_{k_{p}k_{p-1}}^{(p-1)} A_{k'_{p}k'_{p-1}}^{*(p-1)} \\ +     e^{-i x \omega} \
    \left( \sum_{i=1}^{N} M_{i} A_{ik'_{L}}^{(L)} A_{ik_{L}}^{*(L)}\right) A_{k'_{1}1}^{(0)}  A_{k_{1}1}^{*(0)} \prod_{p=2}^{L}  A_{k'_{p}k'_{p-1}}^{(p-1)} A_{k_{p}k_{p-1}}^{*(p-1)}.
\end{multline}
Thus, we generate a Fourier series with coefficients
\begin{equation}
    c_{\omega} = \sum_{\substack{\mathbf{k,k'}=1\\ \Omega_{kk'} = \omega}}^{N} \left( \sum_{i=1}^{N} M_{i} A_{ik_{L}}^{(L)} A_{ik'_{L}}^{*(L)}\right) A_{k_{1}1}^{(0)}  A_{k'_{1}1}^{*(0)} \prod_{p=2}^{L}  A_{k_{p}k_{p-1}}^{(p-1)} A_{k'_{p}k'_{p-1}}^{*(p-1)},
    \label{eq:c_wA1}
\end{equation}
and $c_{-\omega}=c_{\omega}^{*}$, with a frequency spectrum
\begin{equation}
    \Omega_{\textbf{kk'}} = \{\Lambda_{\textbf{k}}- \Lambda_{\textbf{k'}}\} = \{(\lambda_{k_1}+\cdots+\lambda_{k_L})- (\lambda_{k'_1}+\cdots +\lambda_{k'_L})\}.
    \label{eq:freq_app}
\end{equation}
Notice that the frequency spectrum is directly related to the eigenvalues of the encoding Hamiltonian, while the coefficients depend on the elements of the trainable parameters.

\section{Matrix elements combinations to generate Fourier coefficients} \label{app:coeff}

All the terms that constitute a given coefficient with associated frequency $\omega$ have in common that $\Lambda_\textbf{k}- \Lambda_\textbf{k'} = \omega$ (see Eq. \eqref{eq:c_wA1}). In this appendix, we explore how many combinations of the multi-indices $\textbf{k}$, $\textbf{k'}$ give us the same frequency $\omega$, what we call the number of contributions to the coefficient ($s_\omega$). In some occasions, this is also called the degeneracy of the frequency. Having more terms that contribute to the same coefficient may be beneficial since we can have more parameters yielding the same coefficient. The number of contributions does not affect the discussion about the degrees of freedom.

The frequencies are generated according to Eq. \eqref{eq:freq_app}. For one qudit model and with the ``spin-like" encoding discussed in Sec. \ref{sec:FS with quantum circuits}, the frequencies are generated by subtracting $L$ eigenvalues to $L$ eigenvalues, and each of them can take $N$ values. In total we have $|\omega| =N^{2L}$ possible eigenvalues combinations. However, some subtraction results with the same frequency $\omega$. The number of combinations that gives rise to the same frequency is given by
\begin{equation}
    s_{\omega} = \binom{2L}{L-\omega}_{N-1}.
    \label{eq:multiplicity}
\end{equation}
As expected, the number of combinations depends on the number of layers, the frequency considered and the dimension of the model. The symmetry in the generation of the frequencies is reflected by $s_{\omega} = s_{-\omega}$. The sum of all combinations gives back $|\omega|$, the eigenvalues combinations. $\sum_{\omega \in \Omega} s_{\omega} = d^{2L} = |\omega|$. For qubits ($N=2$), the distribution of combinations becomes a binomial distribution:
\begin{equation}
    s_{\omega}^{N=2}= \binom{2L}{L-\omega}_1 = \frac{2L!}{(L-\omega)! (L+\omega)!}.
\end{equation}
For qutrits ($N=3$), we have a trinomial distribution:
\begin{equation}
     s_{\omega}^{N=3} =  \binom{2L}{\omega}_2 = \sum_{\substack{ 0 \leq \mu, \nu \leq 2L \\ \mu + 2\nu = 2L+\omega}} \frac{2L!}{\mu! \nu ! (2L-\mu -\nu)!}.
\end{equation}
For ququarts, a quadrinomial distribution, etc. By using higher-dimensional systems, the number of total combinations grows exponentially with the number of layers used.

For example, let us consider a model with one qubit and $L=2$. The frequency spectrum is given by $\Omega = \{\lambda_{k_1}+ \lambda_{k_2}- (\lambda_{k'_1}+ \lambda_{k'_2})\}$. The eigenvalues are $\lambda_{k_i} = \pm 1/2$ and we can combine them in $|\omega| = d^{2L} = 2^{4}= 16$ ways to obtain one of the frequencies $\Omega = \{-2, -1, 0,1,2\}$. In Tab. \ref{tab_1}, we show the distribution of combinations to obtain these frequencies. As we can see, it fulfils $s_{-2}+ s_{-1}+ s_0 + s_1 + s_2 = 16$.

\begin{table}[t!]
\begin{tabular}{|l|l|l|l}
\cline{1-3}
$\omega = 0$                                            & $\omega = 1$                                            & $\omega = 2$ &  \\ \cline{1-3}
$\frac 12 + \frac 12 - \bigg( \frac 12+ \frac12\bigg)$ &
  $\frac 12 +\frac 12 - \bigg( -\frac 12+\frac12\bigg)$ &
  $\frac 12 +\frac 12 - \bigg( -\frac 12-\frac12\bigg)$ &
   \\ \cline{1-3}
$\frac 12 - \frac 12 - \bigg( \frac 12-\frac12\bigg)$   & $\frac 12 +\frac 12 - \bigg( +\frac 12-\frac12\bigg)$   &              &  \\ \cline{1-3}
$\frac 12 - \frac 12 - \bigg( -\frac 12+\frac12\bigg)$  & $\frac 12 - \frac 12 - \bigg( -\frac 12-\frac12\bigg)$  &              &  \\ \cline{1-3}
$-\frac 12 +\frac 12 - \bigg( -\frac 12+\frac12\bigg)$  & $-\frac 12 + \frac 12 - \bigg( -\frac 12-\frac12\bigg)$ &              &  \\ \cline{1-3}
$-\frac 12 + \frac 12 - \bigg( +\frac 12-\frac12\bigg)$ &                                                         &              &  \\ \cline{1-3}
$-\frac 12 - \frac 12 - \bigg( -\frac 12-\frac12\bigg)$ &                                                         &              &  \\ 
\cline{1-3}
{ $\boldsymbol{s_0 = 6}$} &
  {$\boldsymbol{s_1= 4}$} &
  {$\boldsymbol{s_2= 1}$} &
   \\ \cline{1-3}
\end{tabular}
\caption{Example of the number of combinations of the positive frequencies in the single-qubit model with $2$ layers and encoding gate $S(x) = e^{ixH}$ with $H = \sigma_z/2$. The combinations of the negative frequencies are obtained by swapping the eigenvalues, giving rise to the same combination value ($s_{\omega} = s_{-\omega}$).}
\label{tab_1}
\end{table}

We can easily generalize this for the $M-$dimensional models. The discussion is valid for the Line, Parallel, and Mixed ansatz. We have coefficients associated with $M$ frequencies in such models: $c_{\omega_1,..., \omega_M}$. Hence, the possible combinations of the coefficients depend on the combinations of every single frequency $\omega_i$,
\begin{equation}
    s_{\vec{\omega}} = s_{\omega_1} s_{\omega_2}  ...  s_{\omega_M},
\end{equation}
where $s_{\omega_i}$ with $i\in\{1,...,M\}$ is the degeneracy of the one-dimensional model given in Eq. \eqref{eq:multiplicity}. The more terms are contributing to the sum of a coefficient, the more likely is to exist more than one set of parameters that contributes to the same coefficient. 

\section{Re-scaling factor}\label{app:rescaling}

In this section, we discuss why it is relevant to introduce a re-scaling factor in quantum methods for function fitting, which can be extended to other methods. For simplicity, we first discuss the re-scaling strategy for the one-dimensional case and then generalize it to the multidimensional model. 

The non-negative frequency spectrum generated by the one-qudit models with the ``spin-like" encoding (without the re-scaling factor) is given by
\begin{equation}
    \Omega = \{(d-1)L, (d-1)L-1,...,0\}.
    \label{eq:eigenvalues_general_app}
\end{equation}
Suppose we want to fit a $M-$dimensional function that can be approximated with a Fourier series of degree $D'$. We fit the function with at least $L = D' / (d-1)$ layers, under the condition of the degrees of freedom (see section \ref{sec:Multi FS}). Nevertheless, we can find cases in which there is no number of layers such that $D' = (d-1)L $. For instance, with the ``spin-like" encoding, we generate integer frequencies without the possibility of generating semi-integer or float frequencies, regardless of the number of layers used. If the target function contains a non-integer frequency, the model cannot generate the proper functions and the model's training fails. We overcome this inconvenience by introducing a re-scaling  $\eta$. 

\begin{figure}[t!]                  
\centering
\includegraphics[width=0.6\columnwidth, angle=0]{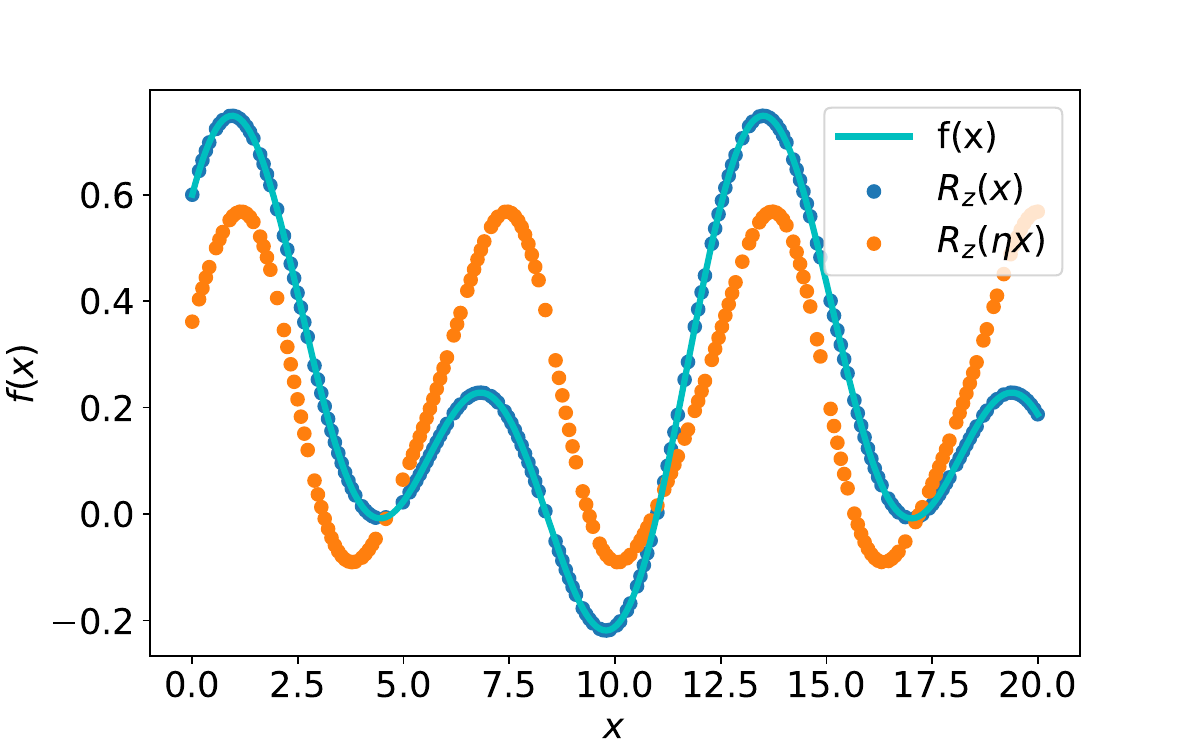}
\caption{Simulations of a one-dimensional Fourier serie with one qubit and two layers. The target function is $f(x) = 0.2(1+\cos{\frac x2}+ \sin{\frac{x}{2}}+ \cos{x} +\sin{x})$ and the trainable gates are $A^{(i)}(\vec{\theta}_{i}) = R_y(\theta^{(1)}_i) R_z(\theta^{(2)}_i)$ in both cases. The solid line is the target function, while the dots correspond to the output of different quantum models. The blue dots use an encoding gate $R_z(x)$ and the orange ones use  $R_z(\eta x)$, where $\eta$ is the re-scaling factor. The optimization subroutine finds $\eta\simeq \frac{1}{2}$.}
\label{fig:re-scaling}
\end{figure}

Let's study the case of fitting a function $f(x) = 0.2(1+\cos{\frac x2}+ \sin{\frac{x}{2}}+ \cos{x} +\sin{x})$, which is decomposed in a Fourier series with frequencies $\Omega' = \{0,  \pm 1/2, \pm 1\}$. With one layer, the qubit model generates the frequency $\omega = 0$ and $\omega = \pm 1$, but it cannot generate the semi-integer ones. Consequently, it fails to fit this function. A solution to avoid this is introducing a re-scaling factor $\eta$ in the Hamiltonian $H$ of the encoding gates, such that $S(x) = e^{ix\eta H} = e^{ixH'}$. Now, all the eigenvalues of the Hamiltonian become $\eta \lambda_i$, leading us to a different frequency spectrum
\begin{equation}
    \Omega_{\eta} = \eta \Omega.
\end{equation}
The re-scaling factor $\eta$ is optimized by the classical subroutine and all the other trainable parameters. In the example considered, the optimization subroutine ideally finds $\eta = \pm \frac{1}{2}$. With this, we change the eigenvalues of the encoding Hamiltonian to $\lambda' = \{-1/4, 1/4 \}$. Consequently, the frequency spectrum becomes $ \Omega_{\eta = 1/2} =  \{\pm L/2, \pm (L-1)/2,...,0\}$. With two layers, we have the desired frequencies present in the target function. A simulation with this example is shown in Fig. \ref{fig:re-scaling}.

To have a more flexible model in terms of the frequencies, one can introduce different re-scaling parameters in all $L$ processing gates to be optimized together with the rest of the model's parameters (see Ref.\cite{DR} for a similar proposal for classification problems). By doing this, we can match the target function frequencies with more freedom. The accessible frequencies for the model become
\begin{equation}
    \Omega_{\vec{\eta}} =\{ \eta_1(\lambda_{k_1}-\lambda_{k'_1})+...+\eta_L( \lambda_{k_L} - \lambda_{k'_L})\}.
    \label{eq:freq_encoding.}
\end{equation}
The Fourier series now has non-equispaced and real frequencies that can capture better the target function structure. We obtain any desired frequency present in the target with enough layers by optimizing the right parameters. 

The re-scaling proposal can be easily generalized to $M-$dimensional models by introducing more re-scaling parameters: $S(\vec{\eta} \cdot \vec{x})$, where $\vec{\eta} = (\eta_1, \eta_2, ...,\eta_M)$ and $\vec{x} = (x_1, x_2,...,x_M)$ are $M-$dimensional vectors. In this way, we modify the frequency associated with each dimension. 

This method requires a small amount of classical pre-processing. Instead of generating a unitary transformation with parameter $\theta = x$, we have to generate a $\theta = \eta_i x$ transformation. In exchange for this, we obtain flexibility to fit functions with unknown frequencies. However, this approach should be treated carefully, because methods with re-scaling parameters may lead to good expressibility but overfitting and poor generalization bounds for more complex tasks.

\section{Fourier series with the Line Ansatz} \label{app:LA}

In this appendix, we explore the Line Ansatz (LA) for a qudit of arbitrary dimension $d$. We show the exact form of the parameterized states after the quantum circuit and the expectation value of an observable in this state (the output of the quantum model). We also include a slight variation of the LA that does not contain the processing gates between each data dimension. We show why this formalism does not fit multidimensional functions in general. 

We study the state after the quantum circuit of the LA model (see Fig. \ref{fig:ansatzes} (\textbf{a}) for the quantum circuit). We consider the ``spin-like" encoding discussed in the main article for all data features, having the same frequency spectrum in all dimensions. The quantum state is given by
\begin{equation}
     \psi_i = \sum_{\substack{j_1,..,j_L \\ k_1,..,k_L  \\t_1,...,t_L\\ \vdots} }^dA^{(L)}_{M_{i t_L}}e^{ix_M \lambda{t_L}}...A^{(L)}_{1_{k_L j_L}}e^{ix_1\lambda{j_L}}...A^{(1)}_{M_{j_2 t_1}}e^{ix_M\lambda{t_1}}...A^{(1)}_{1_{k_1 j_1}}e^{ix_1\lambda{j_1}}A^{(0)}_{j_11},
\end{equation}
where $\lambda_{j_i}$ are the eigenvalues of the single-qudit encoding Hamiltonian $H$, with $i\in \{1,\cdots,d\}$. In the symbol for a trainable gate $A^{(l)}_{m_{i,j}}$, $l$ indicates the layer, $m$ the position of the gate in the layer, and $i,j$ are the indices of the matrices. Note that each eigenvalue on the exponentials, introduced by the encoding gates $S(x_i)$, has a different index. This is crucial for having a non-related dependency in each dimension and it occurs because of the intermediate trainable gates between each encoding gate. For the sake of simplicity, from now on, we assume that the $M$ processing gates of any layer $l$ have the same structure but taking into consideration that they have different parameters: $A_1^{(l)} \sim A_2^{(l)} \sim ...\sim A_M^{(l)} \equiv A^{(l)}$. With the multi-index notation $\textbf{j} = \{j_1,j_2,...,j_L\}$ and re-grouping terms we write
\begin{equation}
    \psi_i= \sum_{\textbf{j}, \textbf{k},...\textbf{t}} ^d e^{i x_1( \lambda_{j_L}+...+\lambda_{j_1})}... e^{i x_M( \lambda_{t_L}+...+\lambda_{t_1})} A^{(L)}_{{i t_L}} ...A^{(L)}_{{k_L j_L}}..A^{(1)}_{{j_2 t_1}}...A^{(1)}_{{k_1 j_1}}A^{(0)}_{j_11}.
\end{equation}
Finally, we define the multi-index sum: $\Lambda_\textbf{j} = \lambda_{j_1}+ \lambda_{j_2}+...+\lambda_{j_L}$, simplifying the previous expression
\begin{equation}
    \psi_i= \sum_{\textbf{j}, \textbf{k},...\textbf{t}}^d e^{i (x_1 \Lambda_{\textbf{j}}+ ...+x_M \Lambda_{\textbf{t}})} A^{(L)}_{{i t_L}} ...A^{(L)}_{{k_L j_L}}..A^{(1)}_{{j_2 t_1}}...A^{(1)}_{{k_1 j_1}}A^{(0)}_{j_11}.
\end{equation}

After this, we compute the expectation value of the observable $\mathcal{M}$. Without loss of generality, we assume that $\mathcal{M}_{ij} = \mathcal{M}_{ii} = \mathcal{M}_i$ 
\begin{equation}
\begin{split}
    \langle \mathcal{M} \rangle = \sum_{\substack{\textbf{j, j'}, ... \textbf{t, t'} \in[N]^{L} \\ \omega_1 = \omega_1',..., \omega_M = \omega_M'}} ^d \sum_{i}^N &e^{ix_1 (\Lambda_\textbf{j}- \Lambda_\textbf{j'})}... e^{ix_M (\Lambda\textbf{t}- \Lambda_\textbf{t'})} A^{(0)*}_{j'_11}A^{(1)*}_{{k'_1 j'_1}}...A^{(1)*}_{{j'_2 t'_1}}...A^{(L)*}_{{k'_L j'_L}}...A^{(L)*}_{{i t'_L}} \mathcal{M}_i \\
    & \times A^{(L)}_{{i t_L}} ...A^{(L)}_{{k_L j_L}}...A^{(1)}_{{j_2 t_1}}...A^{(1)}_{{k_1 j_1}}A^{(0)}_{j_11}.
\end{split}
\end{equation}
As we can observe, the expectation value has the structure of a multidimensional Fourier series: 
\begin{equation}
      \langle \mathcal{M} \rangle =  \sum_{ \Vec{\omega} \in \Vec{\Omega}} c_{\Vec{\omega}} e^{i\Vec{x}\cdot\Vec{\omega}}.
\end{equation}
Each data dimension has its frequency spectrum, resulting in a $M-$dimensional vector of frequencies:
\begin{equation}
    \Vec{\Omega}^{(LA)} = \big(\{\Lambda_{\textbf{j}}-\Lambda_{\textbf{j'}} \} ,...,\{\Lambda_{\textbf{t}}-\Lambda_{\textbf{t'}} \}\big).
\end{equation}
The set of coefficients
\begin{equation}
     c_{\Vec{\omega}} = \sum_{\substack{\Lambda_{\textbf{j}}- \Lambda_{\textbf{j'}} = \omega_1\\\vdots\\ \Lambda_{\textbf{t}}- \Lambda_{\textbf{t'}} = \omega_M}}A^{(0)*}_{j'_11}A^{(1)*}_{{k'_1 j'_1}}...A^{(1)*}_{{j'_2 t'_1}}...A^{(L)*}_{{k'_L j'_L}}...A^{(L)*}_{{i t'_L}} O_i A^{(L)}_{{i t_L}} ...A^{(L)}_{{k_L j_L}}...A^{(1)}_{{j_2 t_1}}...A^{(1)}_{{k_1 j_1}}A^{(0)}_{j_11},
\end{equation}
also fulfill that $c_{\omega_1, \omega_2, \cdots, \omega_M} = c_{-\omega_1, -\omega_2, \cdots, -\omega_M}^*$.

\begin{figure}[t!]                  
\centering
\includegraphics[width=1\columnwidth, angle=0]{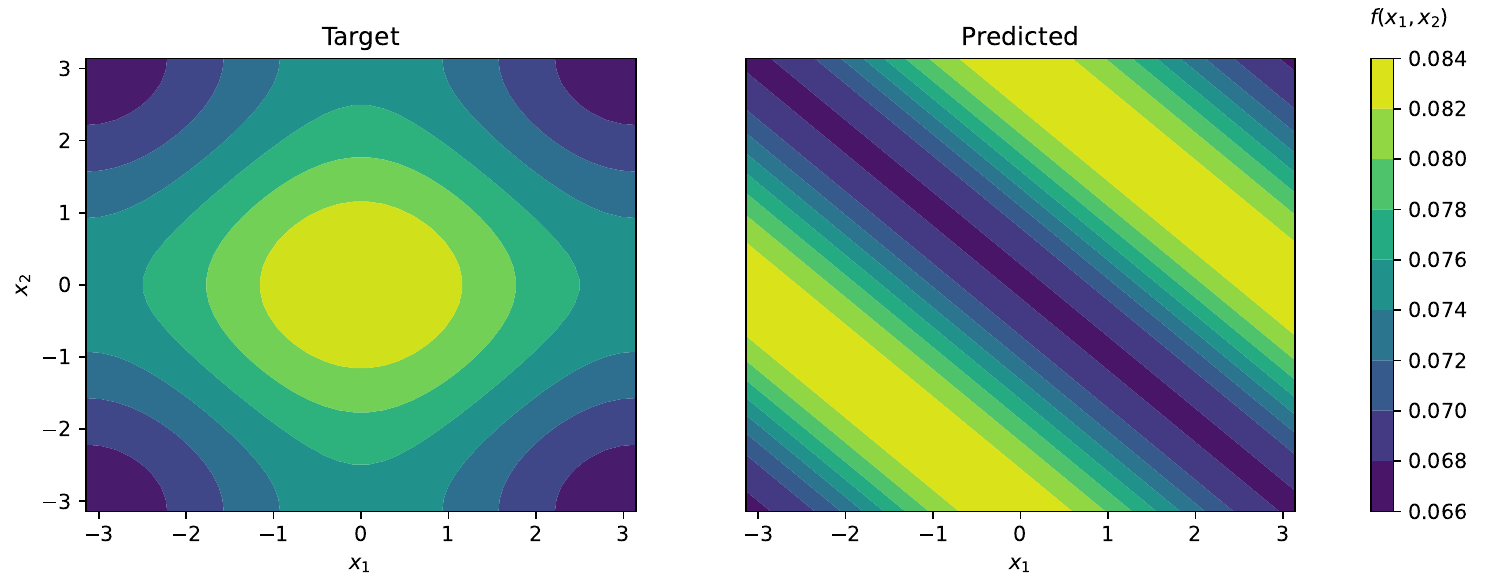}
\caption{Fitting of a Fourier two-dimensional series with $S(\Vec{x}) = R_z(x_1)R_z(x_2)$ and $10$ layers. The target function is $f(x_1, x_2) = \frac{1}{12}(1+ \cos x_1+ \cos x_2)$. The classical optimization method is the Nelder-Mead. We use $500$ training points and $1500$ test points. The predicted functions do not result to be like the target. Instead, it presents a structure similar to $f(x_1+x_2)$.}
\label{fig:Fourier_2d_first}
\end{figure}

Now, we provide an alternative ansatz in which data of all dimensions is uploaded without a processing gate separating the different dimensions. Considering this, the ansatz is composed of the following layers
\begin{equation}
    L_0 = A^{(0)}, \hspace{1.5cm} L_i = S(x_1)S(x_2)...S(x_M) A^{(i)}.
\end{equation}
The problem with this ansatz is that, by using the same $S(x_i)$ for all encoding gates, these gates become $S(x_1)S(x_2)\cdots S(x_M) = S(x_1+x_2+...+x_M)$. Hence, all data dimensions are mapped to a single dimension in the following way: $\Bar{x} = x_1+...+x_M$. In other words, the $M-$dimensional data collapses in a one-dimensional space. For example, the data we introduce for $M=2$ with a qubit circuit is $\bar{x} = x_1+x_2$. The model cannot distinguish the data points $x_1= 1$, $x_2=0$, and $x_1=0$, $x_2=1$. Consequently, the results obtained by the model for fitting a two-dimensional function have a similar structure to the function $f(x_1+x_2)$ (depicted in Fig. \ref{fig:Fourier_2d_first}). For this reason, we need to introduce a trainable gate in the middle of the encoding gate to separate the data dependency.

\section{Fourier series with the Parallel Ansatz} \label{app:PA}

This section explains in more detail the Parallel Ansatz (PA) and studies its performance. With this ansatz, we take advantage of larger Hilbert spaces for operations with the trainable gates, which determine the coefficients of the Fourier series. First, each data dimension is encoded in a different qudit. Then, we present the resulting state after the parameterized quantum circuit and the expectation value of an arbitrary observable $\mathcal{M}$ in this state. We also provide a variation of this ansatz that does not use entanglement in the trainable gates and we discuss the differences between the two models. 

Following the circuit of the PA in Fig. \ref{fig:ansatzes} (\textbf{b}), the $M-$qudit state after the circuit is given by
\begin{equation}
    \psi_i = \sum_{j_1,...,j_L = 1}^{N = d^M}  A^{(L)}_{i j_L}e^{i (x_1 \lambda_{j_L}^{(1)}+ ...+x_M \lambda_{j_L}^{(M)})} ...A^{(1)}_{j_2 j_1}e^{i (x_1 \lambda_{j_1}^{(1)}+ ...+x_M \lambda_{j_1}^{(M)})}A^{(0)}_{j_11}.
    \label{eq:state_PA_A}
\end{equation}
Now the indices run from $1$ to $N = d^M$, being $M=n$, the number of qudits, and $d$ its dimension. The encoding gate is $S (\Vec{x}) = S(x_1)\otimes S(x_2)\otimes...\otimes S(x_M) = e^{ix_1H}\otimes e^{ix_2H}...e^{ix_MH}$, which can be rewritten as $S(\Vec{x}) = e^{i(x_1H^{(1)}+x_2H^{(2)}+...+x_MH^{(M)})}$, with 
\begin{equation}
    H^{(i)}= \underbrace{\mathbb{I}}_1\otimes \underbrace{\mathbb{I}}_2 \otimes \cdots \otimes \underbrace{H}_i \otimes \cdots \otimes \underbrace{\mathbb{I}}_M.
    \label{eq:H_i}
\end{equation}
All the Hamiltonians are built by the tensor product of $n-1$ identities except in the $i$ position where we have the single-qudit Hamiltonian. This gives a Hamiltonian of dimension $N= d^M$, which has the $d$ eigenvalues $\lambda_i$ of the single-qudit Hamiltonian distributed along the $N = d^M$ possible positions on the diagonal. For example, a qubit model with two-dimensional data ($M=2$) and the ``spin-like" encoding ($H = \sigma_z/2$) has the following encoding gate: $S(x_1, x_2) = e^{ix_1H}\otimes e^{ix_2H} = e^{i(x_1 H^{(1)}+x_2 H^{(2)})}$.  The eigenvalues of the two Hamiltonians are: $\lambda^{(1)} = \{ -\frac 12,-\frac 12, +\frac 12, +\frac 12 \}$, and  $\lambda^{(2)} = \{ -\frac 12,+\frac 12, -\frac 12, +\frac 12 \}$. Therefore, the eigenvalues of $\sigma_z/2$ are maintained but distributed differently. Continuing with the quantum state, we group the different eigenvalues with the multi-index sum notation, i.e., $\Lambda^{(i)}_\textbf{j} = \lambda^{(i)}_{j_1}+\lambda^{(i)}_{j_2}+...+\lambda^{(i)}_{j_L}$, we rewrite the final state as
\begin{equation}
    \psi_i = \sum_{\textbf{j}}^{N = d^n} e^{i(x_1\Lambda^{(1)}_\textbf{j}+...+x_M\Lambda^{(M)}_\textbf{j})} A^{(L)}_{i j_L}...A^{(1)}_{j_2 j_1}A^{(0)}_{j_11}.
\end{equation}
All data dimensions in the exponential are multiplied by a multi-index sum with the same index, which may cut some freedom in fitting a Fourier series. Although having the same index, they are non-equally distributed. With this, the model has enough freedom to have an independent frequency spectrum in each dimension. Finally, the expectation value of the observable is given by
\begin{equation}
\begin{split}
    \langle \mathcal{M} \rangle = \sum_{\textbf{j,j'}=1=1}^N\sum_{i=1}^N e^{i[x_1(\Lambda_{\textbf{j}}^{(1)}- \Lambda_{\textbf{j'}}^{(1)})+...+ x_M(\Lambda_{\textbf{j}}^{(M)}- \Lambda_{\textbf{j'}}^{(M)})]}  A^{(0)*}_{j'_1 1} A^{(1)*}_{j'_2 j'_1}...A^{(L)*}_{i j'_L} M_i A^{(L)*}_{i j_L}... A^{(1)}_{j_2 j_1}  A^{(0)}_{j_1 1},
\end{split}
\end{equation}
which has the structure of a multidimensional Fourier series. The observable $\mathcal{M}$ acts on $n$ qudits, but we propose to measure only one qudit, for example, the first one. Then the observable becomes $\mathcal{M} = \mathbf{M} \otimes \mathbb{I}\otimes... \otimes \mathbb{I}$, where $\mathbf{M}$ is a single-qudit observable. This does not affect the models' structure. The frequency spectrum obtained is
\begin{equation}
    \Vec{\Omega}^{(PA)} = (\{\Lambda_{\textbf{j}}^{(1)}- \Lambda_{\textbf{j'}}^{(1)}\},..., \{\Lambda_{\textbf{j}}^{(M)}- \Lambda_{\textbf{j'}}^{(M)}\}).
\end{equation}
Therefore, the PA's output has a multidimensional Fourier series structure.

Now we explore a variation of the PA. The circuit ansatz we consider is the same, but instead of using processing gates $A^{(i)}$ as multi-qudit gates, we only use single-qudit gates, meaning that we need to eliminate any entangling gate. In this way, we explore the outcome of the model working with each qudit separately. Now we measure an observable acting in all qudits because they do not interact and consequently do not share any correlation. The state after the circuit with the ansatz considered is
\begin{equation}
    \ket{\psi} = A^{(0)} S(\vec{x})  A^{(1)}...S(\vec{x})A^{(L)} = \ket{\psi_1}\otimes \ket{\psi_2} \otimes \cdots \otimes \ket{\psi_M}.
\end{equation}
where $A^{(i)} = A^{(i)}_1 \otimes A^{(i)}\otimes... \otimes A^{(i)}_M$, and $S(\vec{x})  = S(x_1)\otimes S(x_2)\otimes... \otimes S(x_M)$.  Hence, if we choose to measure an observable $\mathcal{M} = \sigma_z \otimes \sigma_z \otimes .... \otimes \sigma_z$, the result is 
\begin{equation}
\begin{split}
    \langle \mathcal{M} (\vec{x}) \rangle  = \bra{\psi_1}\sigma_z \ket{\psi_1}\bra{\psi_2}\sigma_z \ket{\psi_2}&...\bra{\psi_M}\sigma_z \ket{\psi_M} = \sum_{\omega_1}c_{\omega_1}^{(1)}e^{ix_1\omega_1}\sum_{\omega_2}c_{\omega_2}^{(2)}e^{ix_1\omega_2}...\sum_{\omega_M}c_{\omega_M}^{(M)}e^{ix_1\omega_M}
    \\ = &\sum_{\vec{\omega}} c_{\omega_1}^{(1)}c_{\omega_2}^{(2)}...c_{\omega_M}^{(M)} e^{i(x_1\omega_1+x_2\omega_2+...+x_M\omega_M)},
\end{split}
\end{equation}
where we have used the results of the one-dimensional Fourier series discussed in App. \ref{app:Fourier}. We obtain a multiplication of $M$ one-dimensional Fourier series. In this case, however, we do not have all the free coefficients because $c_{\vec{\omega}} \neq  c_{\omega_1}^{(1)}c_{\omega_2}^{(2)}...c_{\omega_M}^{(M)}$. For example, for $M=2$ we have $5$ free coefficients: $c_{00}, c_{01}, c_{10}, c_{11}$, $c_{1-1}$ but in this variation of the model we have $4$ free coefficients: $c_{0}^{(1)}, c_0^{(2)}, c_1^{(1)}, c_1^{(2)}$, since the other ones are constrained by $c_{\omega_{i}}^{(i)} = c_{-\omega_{i}}^{(i)*}$. The output of this model is still a Fourier series, but not the most general one. Regardless of this, the possibility of finding applications for some problems is not discarded, as the question remains open.

\section{Fourier series with the Mixed Ansatz} \label{app:MA}

In this section, we explore the last ansatz considered. The circuit in Fig. \ref{fig:ansatzes} (\textbf{c}) is called the Mixed Ansatz (MA) because it combines elements of the LA and the PA: various data dimensions are encoded in the same qudit as in the LA, but also other qudits are used to encode them, like in the PA. The idea of this model is to take advantage of the hardware requirements of the current quantum devices. For instance, if we have a chip with $4$ qubits and $8$ data features, we introduce $2$ dimensions in each qubit.

Let us consider a MA circuit of $p$ qudits and a $M-$dimensional dataset. In the ideal case, we encode $M/p$ data dimensions in every qudit. Of course, $M/p$ will not always be an integer. In such cases, we use fewer encoding gates in the last layer. Since we are encoding different data features in the same qudit, we need extra processing gates between these dimensions, like in the LA. All the trainable gates are $p$-dimensional qudit gates since we encode data in this number of qudits. The quantum state after the circuit is given by
\begin{equation}
\begin{split}
    \psi_i = \sum_{\substack{j_1,..,j_L \\ k_1,..,k_L  \\t_1,...,t_L\\ \vdots} }^{N=d^p} &A^{(L)}_{{M/p}_{i t_L}}e^{i(x_{M-(p-1)}\lambda_{t_L}^{(1)}+ ...+x_M \lambda_{t_L}^{(p)})}...A^{(L)}_{1_{k_L j_L}}e^{i(x_1\lambda_{j_L}^{(1)}+...x_p\lambda_{j_L}^{(p)})}...\\
    & \times A^{(1)}_{{M/p}_{j_2 t_1}}e^{i(x_{M-(p-1)}\lambda_{t_1}^{(1)}+...+x_M\lambda_{t_1}^{(p)})}...A^{(1)}_{1_{k_1 j_1}}e^{i(x_1\lambda_{j_1}^{(1)}+...+ x_p\lambda_{j_1}^{(p)})}A^{(0)}_{j_11}.
\end{split}
\end{equation}
This equation has a different distribution of eigenvalues: $\lambda^{(i)}$ with $i\in \{1,..., p\}$ because we have $p$ qudits. The different distribution comes from the tensor product of the identity in all the $p$ qudits and the encoding Hamiltonian in the $i-$th qudit (see Eq. \eqref{eq:H_i}). By introducing the multi-index and multi-sum notation that we have explored in this work, we reduce the expression to
\begin{equation}
 \psi_i = \sum_{\textbf{j}, \textbf{k}, \textbf{t},...} ^{N=d^p}e^{i(x_1 \Lambda_{\textbf{j}}^{(1)}+...+x_p \Lambda_\textbf{j}^{(p)}+...+ x_{M-(p-1)}\Lambda_{\textbf{t}}^{(1)}+...+x_M\Lambda_{\textbf{t}}^{(p)})}
 A^{(L)}_{{M/p}_{i t_L}}...A^{(L)}_{1_{k_L j_L}}...\\
    A^{(1)}_{{M/p}_{j_2 t_1}}...A^{(1)}_{1_{k_1 j_1}}A^{(0)}_{j_11},
\end{equation}
where $\textbf{j} = \{j_1, j_2,..., j_L\}$ and $\Lambda_{\textbf{j}}^{(i)} = \lambda_{j_1}^{(i)}+\lambda_{j_2}^{(i)}+...+\lambda_{j_L}^{(i)}$. The expectation value of an observable $\mathcal{M}$ in this quantum state is given by
\begin{equation}
\begin{split}
    \langle \mathcal{M} \rangle = \sum_{\textbf{j}, \textbf{j'}, \textbf{k}, \textbf{k'}, \textbf{t},\textbf{t'}...}^{N=d^p} \sum_{i=1}^N e^{i[x_1( \Lambda_{\textbf{j}}^{(1)}- \Lambda_{\textbf{j}'}^{(1)})+...+x_p(\Lambda_\textbf{j}^{(p)}- \Lambda_\textbf{j'}^{(p)})+...+ x_{M-(p-1)}(\Lambda_{\textbf{t}}^{(1)}- \Lambda_{\textbf{t'}}^{(1)})+...+x_M(\Lambda_{\textbf{t}}^{(p)}-\Lambda_{\textbf{t'}}^{(p)})]}
    A^{(0)*}_{j'_11}A^{(1)*}_{1_{k'_1 j'_1}}...\\
    \times A^{(1)*}_{{M/p}_{j'_2 t'_1}}...A^{(L)*}_{1_{k'_L j'_L}}...A^{(L)*}_{{M/p}_{i t'_L}} \mathcal{M}_i
 A^{(L)}_{{M/p}_{i t_L}}...A^{(L)}_{1_{k_L j_L}}...
    A^{(1)}_{{M/p}_{j_2 t_1}}...A^{(1)}_{1_{k_1 j_1}}A^{(0)}_{j_11}.
\end{split}
\end{equation}
In terms of the eigenvalues, we see combined features of the two models: having a different distribution of eigenvalues with the same index (in the data dimensions processed at the same level of depth) and having a different index with the same distribution of eigenvalues (the data-features processed in the same qudit). Therefore, this model also generates a multidimensional Fourier series.

\section{Fourier series with the Super-parallel Ansatz}\label{app:SP}
In this section, we introduce the Super-parallel Ansatz. It contains $n =ML$ qudits. The ansatz in question is depicted in Fig. \ref{fig:ansatzes} (\textbf{d}). It can be seen that now the layers grow in two dimensions: in width and depth. Each layer comprises $L$ encoding blocks, with each block consisting of $M$ single-qudit encoding gates. Similar to the previous ansatz, the Super-parallel Ansatz also generates multidimensional Fourier series, and the derivation is similar to that of the Parallel Ansatz. Thus, we will not provide an explicit derivation here. However, the encoding gates are significantly different in this ansatz. For instance, consider an encoding block of the Super-parallel Ansatz with $L=2$ for $M$-dimensional data:
\begin{equation}
\begin{split}
    S(\vec{x}) =& S(x_1)\otimes...\otimes S(x_M)\otimes S(x_1)\otimes...\otimes S(x_M) = e^{ix_1H}\otimes...\otimes e^{ix_MH} \otimes e^{ix_1H}\otimes...\otimes e^{ix_MH}\\
    &\sum_{j_1 = 1}^{d^{LM}} e^{i(x_1 \lambda_{j_1}^{(1)}+...+ x_M \lambda_{j_1}^{(M)}+x_1 \lambda_{j_1}^{(M+1)}+ ...+x_M \lambda_{j_1}^{(2M)}) } = \sum_{j_1 = 1}^{d^{LM}} e^{i[x_1 (\lambda_{j_1}^{(1)}+\lambda_{j_1}^{(M+1)}) +...+ x_M (\lambda_{j_1}^{(M)}+\lambda_{j_1}^{(2M)})] },
\end{split}
\end{equation}
where $\lambda_{j_1}^{(1)}$ is given in Eq. \eqref{eq:H_i}. We see that each data feature in a single encoding block is multiplied by a sum of $L$ eigenvalues (in this case $L=2$). Consequently, the output Fourier series degree $D$ becomes additionally dependent on the number of layers in depth (the number of encoding blocks for a data feature per layer), yielding $D= (d-1)L^2$. The rest of the derivations can be deduced using a similar approach as in Appendix \ref{app:PA}.

\section{Practical case: fitting 4-dimensional data}

 \begin{figure}[t]                         
\centering
\includegraphics[width=0.75\columnwidth, angle=0]{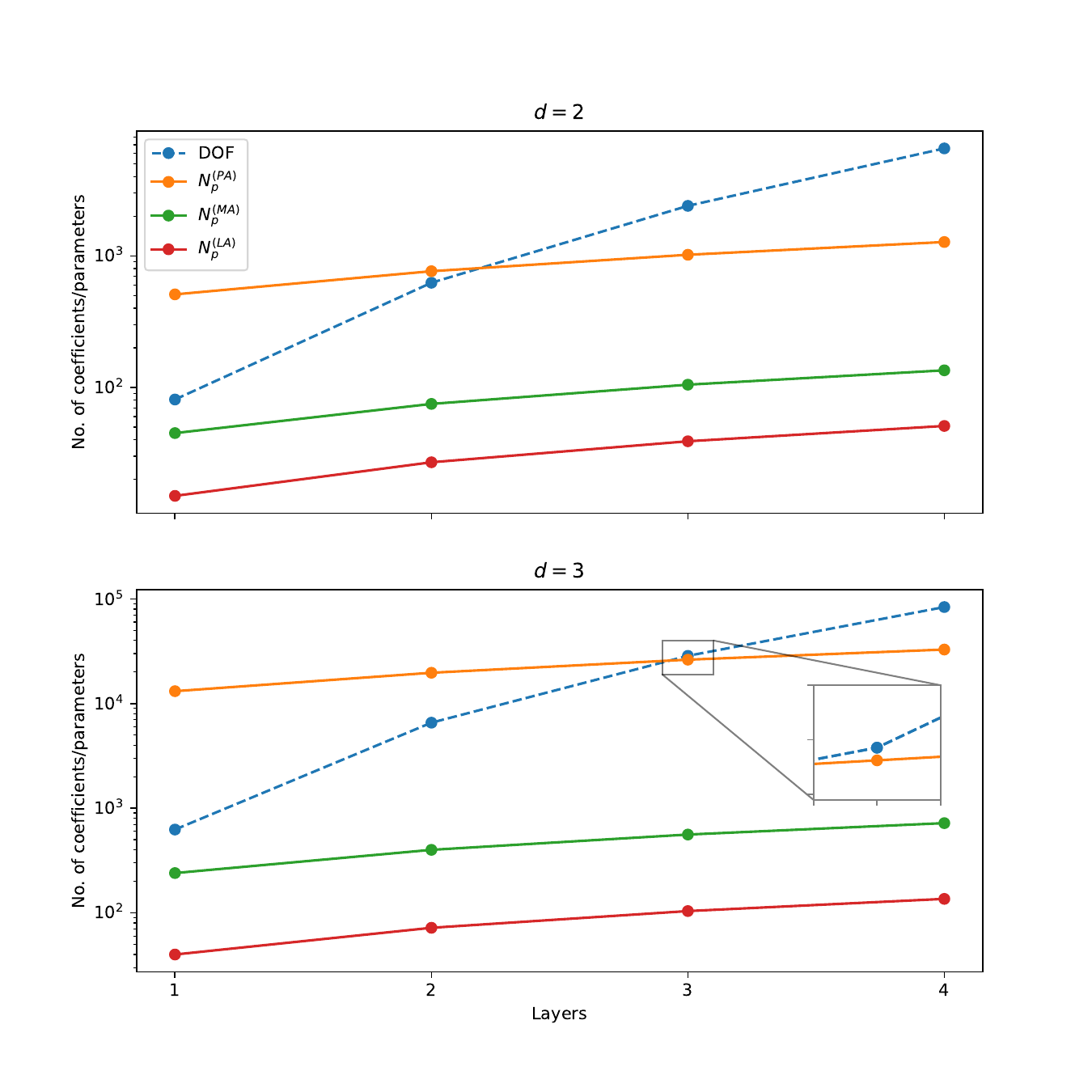}
\caption{Graphic representation of the degrees of freedom and the free parameters with respect to the number of layers for the Line, Parallel, and Mixed ansatzes models with $4-$dimensional data. The upper plot depicts results for qubits and the lower for qutrits. The y-axis is logarithmic and represents the number of coefficients or parameters and the x-axis is the number of layers in the models. The only model that accomplishes the DOF condition is the PA and the results are improved when we use qutrits. With qubits, the condition is satisfied until the Fourier series of degree $D = (d-1)L = 2$ and $D=4$ for qutrits.}
\label{fig:4d}
\end{figure} 

In this appendix, we fit a 4-dimensional Fourier series of degree $D$ with the three ansatzes described in this work: the LA, PA, and MA. We compare which models are more convenient for fitting higher-degree Fourier series. The general target function is given by 
\begin{equation}
    f(x_1, x_2, x_3, x_4) = f(\Vec{x}) =  \sum_{\Vec{\omega} =- D}^{D} c_{\Vec{\omega}} e^{i \Vec{x} \cdot \Vec{\omega}}, 
\end{equation}
where $\Vec{\omega} = (\omega_1, \omega_2, \omega_3, \omega_4)$ describes the frequency in each dimension. The number of free coefficients $N_c$ in the 4-dimensional Fourier series is determined by the degree of the series:
\begin{equation}
    N_c = \frac{(2D+1)^4-1}{2}+1,
\end{equation}
since half of them are constrained by $c_{\Vec{\omega}} =c_{-\Vec{\omega}}^*$, except $c_{\Vec{0}}$. At the same time, the degree of the Fourier series generated by the model is determined by the qudit used and the number of layers of the model: $D = (d-1)L$. Hence, $N_c$ depends on $d$, and $L$. We know that $\nu \equiv 2N_c-1$ is the number of degrees of freedom required for the model to fit the general Fourier series. The number of free parameters of the different models for $4-$dimensional data is given by
\begin{equation}
\begin{split}
    N_p^{(LA)} &= (4L+1)(d^2-1),\\
    N_p^{(PA)} &= (L+1)(d^8-1),\\
    N_p^{(MA)} &= (2L+1)(d^4-1),
\end{split}
\end{equation}
where for the MA we use two qudits, therefore $p=2$. With this, we determine graphically the Fourier series degree that accomplishes the DOF condition. This is shown in Fig. \ref{fig:4d} for qubit and qutrit models. As we can notice, for $4-$dimensional data, both the LA nor the MA do not accomplish the DOF condition for any number of layers. This means we do not have enough free parameters in the model for fitting an arbitrary Fourier series. With the PA model, the condition is satisfied until $L=2$, meaning that we achieve Fourier series of degree $D= 2(d-1)$, having degree $2$ for the qubit model and $4$ for the qutrit. Hence, this model can give limited approximations to some functions.
\end{document}